\documentclass[aps,pre,longbibliography,twocolumn,showpacs,noeprint]{revtex4-2}

\usepackage{CJK}

\usepackage{graphicx}

\usepackage{amsmath}
\usepackage{dcolumn}

\usepackage{bm}
\usepackage[normalem]{ulem}
\usepackage{color}
\usepackage{hyperref}
\usepackage[ruled, vlined, linesnumbered]{algorithm2e}

\usepackage{floatrow}
\usepackage[caption=false]{subfig}

\usepackage{epstopdf}
\newcommand{\bew}{\begin{widetext}}
\newcommand{\ew}{\end{widetext}}

\newcommand{\ii}{{\rm i}}

\newcommand{\bp}{\mathbf{p}}
\newcommand{\bq}{\mathbf{q}}

\newcommand{\br}{\mathbf{r}}

\newcommand{\bff}{\mathbf{f}}
\newcommand{\bu}{\mathbf{u}}

\newcommand{\bh}{\mathbf{h}}
\newcommand{\bg}{\mathbf{g}}

\newcommand{\sep}{ \ \ \ , \ \ \ }

\newcommand{\beq}{\begin{equation}}
\newcommand{\eeq}{\end{equation}}
\newcommand{\beqn}{\begin{eqnarray}}
\newcommand{\eeqn}{\end{eqnarray}}
\newcommand{\pp}{\partial}
\newcommand{\dd}{{\rm d}}
\newcommand{\ee}{{\rm e}}

\newcommand{\la}{\langle}

\newcommand{\ra}{\rangle}

\newcommand{\vnab}{{\bf \nabla}}

\newcommand{\tr}{{\rm Tr}}
\newcommand{\bP}{{\bf P}}
\newcommand{\bF}{{\bf F}}
\newcommand{\bG}{{\bf G}}
\newcommand{\bK}{{\bf K}}

\newcommand{\bC}{{\bf C}}
\newcommand{\bS}{{\bf S}}
\newcommand{\id}{{\rm \bf id}}

\newcommand{\Gr}{{\rm Gr}}

\newcommand{\diagram}[1]{\begin{array}{l}\includegraphics[page=#1]{diagrams.pdf}\end{array}}

\begin{document}
\title{Novel critical phenomena in compressible polar active fluids: Dynamical and Functional Renormalization Group Studies}
\author{Patrick Jentsch}
\email{p.jentsch20@imperial.ac.uk}
\address{Department of Bioengineering, Imperial College London, South Kensington Campus, London SW7 2AZ, U.K.}
\author{Chiu Fan Lee}
\email{c.lee@imperial.ac.uk}
\address{Department of Bioengineering, Imperial College London, South Kensington Campus, London SW7 2AZ, U.K.}
\date{\today}

	\begin{abstract}
Active matter is not only relevant to living matter and diverse nonequilibrium systems, but also
constitutes a fertile ground for novel physics. Indeed, dynamic renormalization group (DRG) analyses have uncovered many new universality classes (UCs) in polar active fluids (PAFs) - an archetype of
active matter systems. However, due to the inherent technical difficulties in the DRG methodology, almost all previous studies have been restricted to polar active fluids in the incompressible or infinitely compressible (i.e., Malthusian) limits, and, when the $\epsilon$-expansion was used in conjunction, to
the one-loop level. Here, we use functional renormalization group (FRG) methods to bypass some of these
difficulties and unveil for the first time novel critical behavior in compressible polar active fluids,
and calculate the corresponding critical exponents beyond the one-loop level. 
Specifically, we investigate the multicritical point of compressible PAFs, where the critical order-disorder transition coincides with critical phase separation. 
We first study the critical phenomenon using a DRG analysis and find that it is insufficient since two-loop effects are important to obtain a nontrivial correction to the scaling exponents. We then remedy this defect by using a FRG analysis. We find three novel universality classes and obtain their critical exponents, which we then use to show that at least two of these universality classes are out of equilibrium because they violate the fluctuation-dissipation relation.

	\end{abstract}
	
\maketitle

\section{Introduction}

Active matter refers to many-body systems in which the microscopic constituents can exert forces or stresses on their surroundings and, as such, detailed balance is broken at the microscopic level \cite{ramaswamy_annrev10,marchetti_rmp13}. However, even if the microscopic dynamics are fundamentally different from more traditional systems considered in physics, it remains unclear whether novel behavior will emerge in the hydrodynamic limits (i.e., the long time and large distance limits \cite{anderson_science72}). One unambiguous way to settle this question is to identify whether the system's dynamical and temporal statistics are governed by a new
universality class (UC), typically characterized by a set of scaling exponents \cite{hohenberg_rmp77,goldenfeld_b92,cardy_b96}. 
These exponents can in principle be determined using either simulation or renormalization group (RG) methods. However, simulation studies can be severely plagued by finite-size effects (e.g., two recent controversies concern the scaling behavior of active polymer networks \cite{sheinman_prl15,pruessner_prl16} and critical motility-induced phase separation \cite{siebert_pre18,partridge_prl19,maggi_softmatt21}). Therefore, RG analyses remain as of today the gold standard in the categorization of dynamical systems into distinct UCs. This perspective has been particularly fruitful in biological physics, where many new nonequilibrium universality classes have been discovered in biology inspired systems \cite{gelimson_prl15,caballero_jsm18,mahdisoltani_prr21,vanderkolk_a22}. Specifically, for polar active fluids (PAFs) \cite{vicsek_prl95,toner_prl95,toner_pre98}, an archetype of active matter systems, the use of dynamic renormalization group (DRG) \cite{forster_pra77} analyses have led to, on one hand, surprising realizations that certain types of PAFs are no different from thermal systems in the hydrodynamic limit \cite{chen_natcomm16, chen_pre18}, and on the other hand 
discoveries of diverse novel phases \cite{toner_prl95,toner_pre98,toner_prl12,toner_prl18,toner_pre18,chen_njp18,chen_prl20,chen_pre20,chen_a22a,chen_pre22,chen_a22b}, critical phenomena \cite{chen_njp15,cavagna_a21,zinati_a22} and discontinuous phase transitions \cite{dicarlo_njp22}. However, due to the inherent technical difficulties in DRG methods, all of these studies have been restricted to PAFs in the incompressible or infinitely compressible (i.e., Malthusian) limits except for rare exceptions
\cite{toner_prl18,toner_pre18}. Further, when a DRG analysis was used in conjunction with the $\epsilon$-expansion method, which was typically the case, it has always been restricted to the one-loop level. 

In this work, we apply for the first time functional RG (FRG) methods on {\it compressible} PAFs and overcome some of these technical challenges. Specifically, we investigate a multicritical region of dry compressible PAFs. Although experimentally less accessible than simple critical points, multicritical points (MCPs) can offer surprising new physics, even in models that are thought to be well understood. For instance, nonperturbative fixed points have been discovered in the extensively studied $O(N)$ model \cite{yabunaka_prl17}, and in systems where two order parameters compete, whose individual critical points belong to equilibrium universality classes, the multicritical region where both critical points coincide can be manifestly out of equilibrium and demonstrate very interesting, spiral phase diagrams \cite{young_prx20}.

We will first apply a traditional one-loop DRG approach to the MCP of our interest, demonstrating how it is insufficient to capture its universal physics and then, for the first time for PAFs, apply a FRG  \cite{wetterich_plb93,morris_ijmpa94,ellwanger_zfpc94,berges_pr02,kopietz_b10,delamotte_lnip12,dupuis_pr20,canet_jopa11} analysis that goes beyond the equivalent perturbative one-loop level.

FRG analyses are intrinsically non-perturbative and are based on an {\it exact} RG flow equation to which approximate solutions can be readily obtained numerically. Recent successes in the applications of FRG include the elucidation of scaling behavior in, e.g., critical and multicritical $N$-component ferromagnets \cite{depolsi_pre20, eichhorn_pre13,boettcher_pre15, yabunaka_prl17}, reaction-diffusion systems \cite{canet_prl04a,canet_prl04b,buchhold_pre16,canet_prl05,tarpin_pre17}, the Kardar-Parisi-Zhang model \cite{canet_prl10,canet_pre11,mathey_pre17}, and turbulence \cite{tomassini_PLB97,mejia-monasterio_pre12,canet_pre16,canet_pre17,pagani_pof21}, as well as non-universal observables far from scaling regimes \cite{daviet_prl19,jentsch_prd22}. Using FRG, we uncover here three novel nonequilibrium UCs by studying a multicritical region of dry compressible PAFs and quantify the associate scaling behaviors {\it beyond} the one-loop level. 

The outline of this paper is as follows. In Sec.~\ref{sec:paf}, we introduce the hydrodynamic theory of  compressible polar active matter and discuss salient features in its phase diagram, which enables us to define the multicritical point of interest. We then show how general scaling invariance of the equations of motion leads to powerlaw behavior in the correlation functions in Sec.~\ref{sec:scale_inv}. For the multicritical point, we show this first in the linear regime in Sec.~\ref{sec:linear} and then turn to the nonlinear regime in Sec.~\ref{sec:nonlin}. In Sec.~\ref{sec:drg}, we perform the one-loop DRG calculation and argue why it is not sufficient to take into account the nonlinearities and then present the FRG approach in Sec.~\ref{sec:frg} that we use instead. Using this method we find three RG fixed points which represent three novel nonequilibrium UCs. We discuss them and their scaling behavior in Sec.~\ref{sec:fps}. Finally, we summarize our findings and give an outlook on future work in Sec.~\ref{sec:summary}.

\section{Compressible polar active fluids}
\label{sec:paf}

\subsection{Equations of motion from symmetry and conservation laws}

Polar active matter aims to describe the collective behavior of swarming animals, e.g., flocks of birds, schools of fish or bacterial swarms. 
In the fluid state, as in passive fluids that  are describable by the Navier-Stokes equations, 
the relevant dynamical variables are the momentum density and density fields, denoted by $\bg$ and $\rho$ respectively.
Without any assumptions about the microscopic realization, one can then, based on symmetry and conservation laws, construct a generic set of hydrodynamic equations of motion (EOM) for these variables. 
Here, we assume that the particle number is conserved (as opposed to, e.g., a Malthusian system in which birth and death of particles can occur \cite{toner_prl12, chen_prl20, chen_pre20}). We thus arrive at a continuity equation as the EOM of the density field $\rho$:
\beq
\label{eq:cont}
\pp_t \rho +\vnab \cdot \bg =0\ .
\eeq 
\\

For the momentum density field, we assume temporal, translational, rotational and chiral invariance. In addition, we focused on active systems in which the constituents move on a fixed frictional substrate, i.e., {\it dry} active matter systems (as opposed to {\it wet} active systems such as active suspensions) \cite{marchetti_rmp13}.
This symmetry consideration leads to the following generic hydrodynamic EOM of $\bg$ \cite{toner_prl95, toner_pre98, toner_pre12}:
\bew
\beq
\label{eq:TT}
\pp_t \bg +\lambda_1 \vnab ( | \bg |^2) +\lambda_2 (\bg \cdot \vnab) \bg +\lambda_3 \bg (\vnab \cdot \bg)
= \mu_1 \nabla^2 \bg +\mu_2\vnab (\vnab \cdot \bg) -\alpha \bg -\beta |\bg|^2\bg -\kappa \vnab \rho +...+\bff 
\ ,
\eeq
\ew
where the ellipsis represents the omitted higher-order terms (i.e. terms of higher order in spatial derivatives and the momentum density field).

In the EOM above, all coefficients are functions of $\rho$, and the noise term $\bff(\br,t)$ is a zero mean Gaussian white noise of the form
\beq
\label{eq:noise_def}
\langle f_i(\br,t)f_j(\br',t')\rangle=
2D\delta_{ij}\delta ^d(\br-\br')\delta(t-t')\,. 
\eeq

The above hydrodynamic EOM (\ref{eq:cont},\ref{eq:TT}) are termed the Toner-Tu EOM \cite{toner_prl95,toner_pre98}. However, in contrast to the original Toner-Tu formulation, we have chosen to use the momentum field as the hydrodynamic variable instead of the velocity field so that there is a linear relationship between $\rho$ and $\bg$, which facilitates our discussion later.

\subsection {Mean-field theory and homogeneous phases}

Given the hydrodynamic EOM, one of the first, and simplest, question to ask is: what are the mean-field homogeneous solutions to the EOM? Answering this question amounts to focusing on temporally invariant, spatially homogeneous, and noise-free solutions to the EOM. These solutions are readily seen to be
\beq
|\bg| = \left\{
\begin{array}{ll}
\sqrt{\frac{|\alpha|}{\beta}}\ , \ & {\rm if} \ \alpha<0 
\\
0\ , &{\rm otherwise}\ ,
\end{array}
\right. 
\eeq
where $\beta$ is taken to be always positive for reasons of stability.
Since $\bg$ can point in any direction, the $|\bg|>0$ state implies  a spontaneous symmetry breaking of the rotational symmetry, and corresponds to the homogeneous ordered phase of polar active matter where collective motion emerges. In contrast, the $|\bg|=0$ state corresponds to the homogeneous disordered phase, i.e., there is no collective motion.

\subsection{Phase diagram: phase separations, critical and multicritical phenomena}

\begin{figure}
		\begin{center}
		\includegraphics[width=\columnwidth]{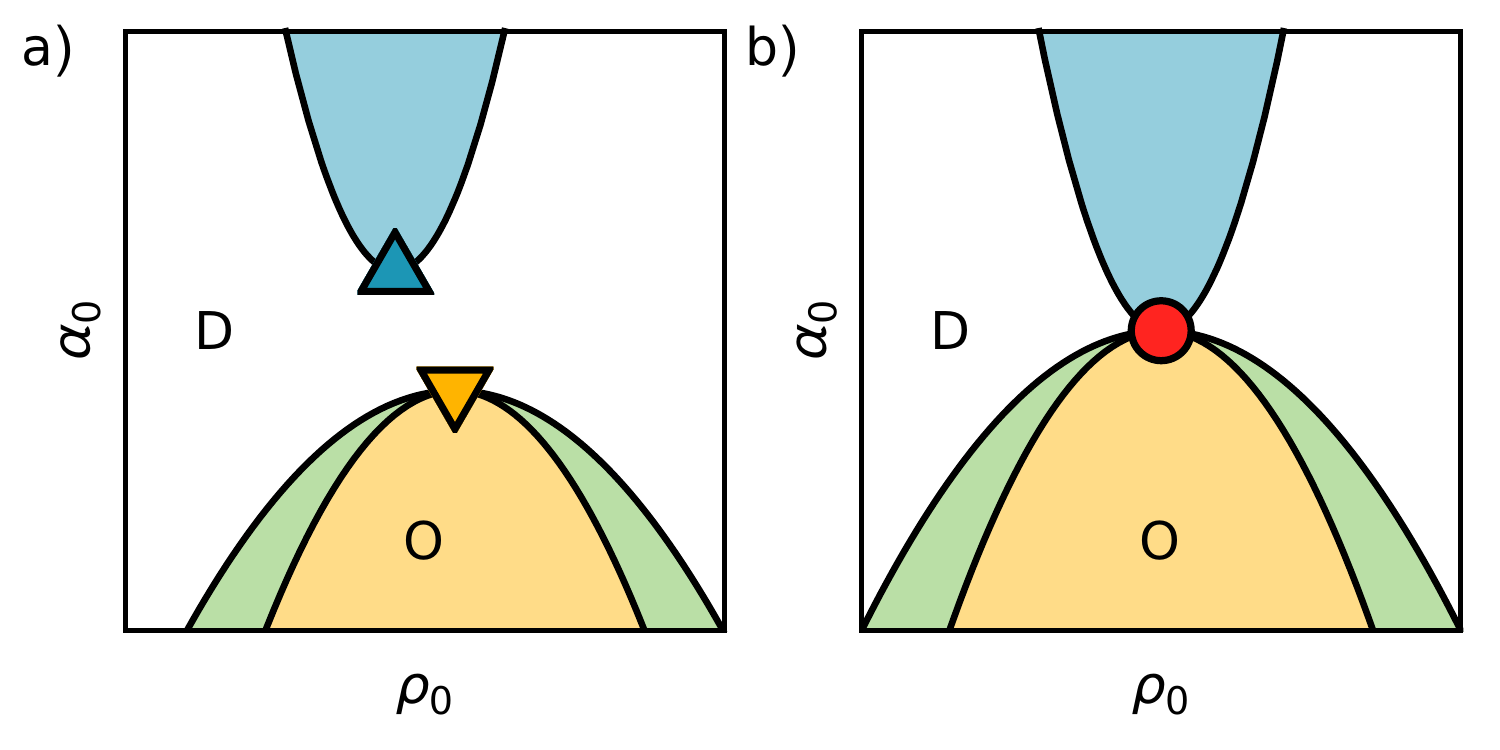}
		\end{center}
		\caption{
		Polar active fluids admit diverse phase transitions and phase separations. 
These figures show qualitatively two possible instances already discussed in \cite{bertrand_a20}. 
		(a) Depending on the model parameter, e.g., ${\alpha_0}$, and the average density $\rho_{0}$, the system can be in the homogeneous disordered phase (white region denoted by D) or the polar ordered phase (yellow region denoted by O); It can also
		phase separate into two disordered phases with different densities (blue region), or into one ordered phase and one disordered phase, again with different densities (green region flanking the homogeneous ordered phase). The critical behavior associated with the first type of phase separation is generically described by the Ising UC (blue triangle) \cite{partridge_prl19}, and that associated with the second type is described by a putatively novel UC yet to be described (yellow inverted triangle) \cite{nesbitt_njp21}. (b) Upon further fine-tuning, these two critical points can coincide (red circle) \cite{bertrand_a20}, and the resulting critical point is described by a novel UC uncovered in the present work.} 
		\label{fig:cartoon}
	\end{figure}

The homogeneous phases discerned from the previous mean-field analysis are, however, not always stable, even in the absence of the noise term $\bff$ (\ref{eq:noise_def}). The standard way to ascertain the in/stability of the homogeneous phases in this noiseless regime is to use a linear stability analysis. Here, the temporal evolution of an initially small perturbation to a homogeneous solution is studied and the growth or decay of its amplitude signifies whether the homogeneous state is unstable or stable, respectively. Since the nature of the inhomogeneous states does not follow from linear stability analysis alone and inhomogoneous, analytic solutions of the noiseless mean-field equations are most often very difficult, it is typically explored via simulations. Using this method, complex phase diagrams of polar active fluids have been uncovered  \cite{bertin_JoPA09,nesbitt_njp21,bertrand_a20}. In particular, distinct types of bulk phase separations, i.e., an inhomogeneous state where two different phases co-exist, have been demonstrated. As the large length and time-scale-limit of all these different models, it is expected that the hydrodynamic EOM captures the same phenomenology in an encompassing phase diagram, schematically depicted in Fig. \ref{fig:cartoon}.

In particular, expressing $\alpha$ and $\kappa$ in Eq.~(\ref{eq:TT}) as 
\beq
\alpha = \sum_{n\geq 0} \alpha_n \delta \rho^n
\sep
\kappa = \sum_{n \geq 0} \kappa_n \delta \rho^n
\ ,
\eeq
where $\delta \rho = \rho -\rho_0$ with $\rho_0$ being the average particle density in the system, two disordered phases (with distinct densities) can co-exist if $\alpha_0>0$ and $\kappa_0<0$ (blue region in Fig.~\ref{fig:cartoon}(a)) \cite{partridge_prl19},
while an ordered phase can co-exist with a disordered phase if $\kappa_0>0$ and $\alpha_0<0$ (green region)\cite{nesbitt_njp21}. Further, the system can become critical upon fine-tuning: if $\alpha_0>0$ and $\kappa_0=\kappa_1=0$, the resulting critical behavior belongs to the Ising universality class (UC) (blue  triangle) \cite{partridge_prl19}, while if $\alpha_0=\alpha_1=0$ and $\kappa_0>0$, the associate critical behavior corresponds to a yet to be characterized UC (yellow  inverted triangle) \cite{nesbitt_njp21}. Recently, a third type of critical behavior 
was identified \cite{bertrand_a20}, which corresponds to the merging of these two distinct critical points by simultaneously fine-tuning $\alpha_0$, $\alpha_1$, $\kappa_0$ and $\kappa_1$ to zero (red circle in Fig.~\ref{fig:cartoon}(b)). {\it The universal behavior of this new multicritical point is the focus of this paper.} 
 
It is interesting to note that apart from these, either homogeneous or bulk phase-separated, states that join at the MCP, different states of microphase separation have been observed as well \cite{solon_prl15, tjhung_prx18, shi_prl20}.

\section{Scale invariant equations of motion}
\label{sec:scale_inv}
It is generally expected that the EOM of general systems at critical points become invariant under rescaling of lengths, time and fields \cite{hohenberg_rmp77,goldenfeld_b92, cardy_b96}. Hence, at the multicritical point (MCP), we expect that the EOM (\ref{eq:cont},\ref{eq:TT}) are invariant under the rescaling
\beq
\label{eq:rescale}
\br \to \br\ee^{\ell},~~t\to t \ee^{z\ell},~~ \rho \to \rho \ee^{\chi_\rho \ell},~~\bg \to \bg \ee^{\chi_g \ell}
\,,
\eeq
for some exponents $z$, $\chi_\rho$ and $\chi_g$, that are a priori not known. If this is the case, however, this defines a rescaling symmetry of the theory which the correlation functions have to obey as well. Take for example the density-density correlation function:
\beq
C_{\rho}(\br,t)=\la \rho(\br, t) \rho({\bf 0},0) \ra = \ee^{ -2\chi_\rho \ell} \la \rho(\br\ee^\ell, t\ee^{z \ell}) \rho({\bf 0},0) \ra  \ .
 \eeq
Choosing $\ell = -\ln r$, $r$ being measured against some reference scale, we see immediately that
\beq
\label{eq:corr_rho}
C_{\rho}(\br,t) = r^{2 \chi_\rho} S_{\rho\rho} \left(\frac{t}{r^{z}} \right)\ ,
\eeq
where $S_{\rho\rho}(.)$ is a scaling function that only depends on the ratio ${t}/{r^{z}}$ which is invariant under rescaling (\ref{eq:rescale}).
So we immediately see that, if the EOM are invariant under a rescaling transformation, the correlation functions will generally express powerlaw behavior.

Likewise, this argument can be applied to the momentum-momentum correlation function
\beq
\label{eq:corr_g}
\bC_g(\br,t) = \la \bg(\br, t)  \bg ({\bf 0},0) \ra =r^{2\chi_g} \bS_{gg}\left(\frac{t}{r^{z}}\right)
\ ,
 \eeq
where $\bS_{gg}$ is again a scaling function with similar properties as $S_{\rho\rho}$.

Ultimately, we will demonstrate that the EOM does become scale invariant and determine the scaling exponents using a FRG analysis, but first, we will illustrate the scale invariance discussed here using the simple, but quantitatively incorrect, linear theory.

\section{Linear regime}
\label{sec:linear}
In the linear regime, i.e., when the non-linear terms in Eq.~(\ref{eq:TT}) are neglected, the scaling behavior discussed above can readily be seen.
Around the  MCP at which critical disordered phase separation (blue triangle in Fig.~\ref{fig:cartoon}a)) merges with critical disorder-order of a generic compressible PAF (yellow inverted triangle),
$|\bg|\approx 0$, such that the linearized EOM are 
\begin{subequations}
\label{eq:lineom}
\begin{align}
\pp_t \rho &=-\vnab \cdot \bg 
\ ,
\\
\pp_t \bg&=\ \mu_1 \nabla^2 \bg +\mu_2\vnab (\vnab \cdot \bg)+\zeta \nabla^2\nabla \rho+\bff 
\ ,
\end{align}
\end{subequations}
where we have introduced the term characterized by $\zeta$ since, when $\kappa_0$ is fine-tuned to zero, this term is now the leading order term linear in $\rho$.
In Eq.~(\ref{eq:lineom}), we have redefined $\rho$ to be $\delta \rho$ to ease notation, and we will continue to do so from now on.

\subsection{Scaling exponents}
\label{sec:linscaling}

\subsubsection{ Correlation functions}

Upon rescaling time, lengths, and fields according to Eq.~(\ref{eq:rescale}), the linearized EOM (\ref{eq:lineom}) become
\begin{subequations}
\begin{align}
\ee^{(\chi_\rho-z)\ell} \pp_t \rho &=-\ee^{(\chi_g-1)\ell}\vnab \cdot \bg 
\ ,
\\
\nonumber
\ee^{(\chi_g-z)\ell} \pp_t \bg&=\ee^{(\chi_g-2)\ell} \left[\mu_1 \nabla^2 \bg +\mu_2\vnab (\vnab \cdot \bg)\right]
\\
&\ \ \ +\ee^{(\chi_\rho-3)\ell} \zeta \nabla^2\nabla \rho+\ee^{-(z+d)\ell/2}\bff 
\ .
\end{align}
\end{subequations}
They thus remain unchanged if
\begin{equation}
	\label{eq:linexp}
	z^{\rm lin}=2,~~\chi_\rho^{\rm lin}=\frac{4-d}{2},~~\chi_g^{\rm lin} =\frac{2-d}{2} \ .
\end{equation}

At the linear level we can therefore directly conclude that
\begin{subequations}
\begin{align}
C_{\rho}(\br,t) &= r^{2 \chi_\rho^{\rm lin}} S^{\rm lin}_{\rho\rho} \left(\frac{t}{r^{z^{\rm lin}}} \right)
\ ,
\\
 \bC_g(\br,t) &=r^{2\chi_g^{{\rm lin}}} \bS^{\rm lin}_{gg}\left(\frac{t}{r^{z^{\rm lin}}}\right)
\ , 
\end{align}
\end{subequations}
using the argument from Sec.~\ref{sec:scale_inv}. 

Since the linearized EOM (\ref{eq:lineom}) are  solvable analytically by performing a spatio-temporal Fourier transform, the scaling behavior of the correlation functions  can in fact be demonstrated explicitly. This has the added advantage that the expressions of the aforementioned scaling functions, $S_{\rho\rho}$ and $\bS_{gg}$, can be obtained in the form of integrals.

Specifically, 
by performing a spatiotemporal Fourier transform, the linear EOM can be written as
\begin{subequations}
\label{eq:linear_sol}
\begin{align}
\rho(\tilde{\bq}) &= \frac{q}{\omega}G_\parallel(\tilde{\bq}) f_\parallel (\tilde{\bq}) \ ,
\\
g_\parallel(\tilde{\bq}) &= G_{\parallel}(\tilde{\bq}) f_\parallel (\tilde{\bq})\ ,
\\
\bg_\perp(\tilde{\bq}) &= G_{\perp}(\tilde{\bq}) \bff_\perp (\tilde{\bq})\ ,
\end{align}
\end{subequations}
where $g_\parallel(\tilde{\bq}) = \bg(\tilde{\bq}) \cdot \hat{\bq}$, $\bg_\perp = \bg - g_\parallel \hat{\bq}$, with $\hat \bq$ being the unit vector in the direction of $\bq$, $q=|\bq|$, $\tilde q=(\bq ,\omega)$, and the $G$'s in (\ref{eq:linear_sol}), or the ``propagators", are:
\begin{subequations}
\label{eq:propagators}
\begin{align}
G_\parallel(\tilde{\bq}) &=  \frac{ \omega}{-\ii(\omega^2-\zeta q^4) +\omega \mu_\parallel q^2} \ ,
\\
G_\perp(\tilde{\bq}) &=  \frac{1}{-\ii\omega +\mu_1 q^2}
\ ,
\end{align}
\end{subequations}
where $\mu_\parallel \equiv \mu_1+\mu_2$.

Given the above expressions, the correlation functions are now obtained straightforwardly:
\begin{subequations}
\label{eq:lin_response}
\begin{align}
	\label{eq:corr_lin_rho}
    C_\rho(\br,t)&=\int_{\tilde \bq} e^{\ii\tilde\bq\cdot\tilde \br} \frac{2Dq^2}{(\omega^2-\zeta q^4)^2 +\omega^2 \mu_\parallel^2 q^4}
     \ , 
    \\
    \bC_g&=\bC_g^\bot+\bC_g^\parallel \ ,
    \\
    \bC_g^\bot(\br,t) &=\int_{\tilde \bq} e^{\ii\tilde\bq\cdot\tilde \br} \frac{2D \bP^\bot(\bq)}{ \omega^2+\mu_1^2 q^4}\ , 
    \\
	\label{eq:corr_lin_g_para}
    \bC_g^\parallel(\br,t)&= \int_{\tilde \bq} e^{\ii\tilde\bq\cdot\tilde \br} \frac{2D \omega^2 \bP^\parallel(\bq)}{ (\omega^2-\zeta q^4)^2+\omega^2\mu_\parallel^2 q^4} \ ,
\end{align}
\end{subequations}
where $\int_{\tilde\bq} \equiv \int \dd^d \bq \dd \omega /(2\pi)^{(d+1)}$,
$\tilde\bq\cdot\tilde \br = \bq\cdot\br-\omega t$, and $P_{ij}^\parallel(\bq)\equiv q_i q_j/q^2$ and $P_{ij}^\perp(\bq)\equiv \delta_{ij} - q_i  q_j/ q^2$ are the projectors parallel and transverse to $\bq$ respectively.

Focusing on $C_\rho(\br,t)$ (\ref{eq:lin_response}a) as an example, the substitutions $\omega = \Omega/r^2$ and $\bq = {\bf Q}/r$ in the integral lead to
\beq
C_\rho(\br,t)= r^{ 4-d}  \int  \frac{\dd^d {\bf Q} \dd \Omega}{(2\pi)^{(d+1)}} \frac{2D Q^2 \exp\left[\ii\left(  {\bf Q}\cdot \hat{\br}-\frac{\Omega t}{r^2} \right)  \right]}{ (\Omega^2-\zeta Q^4)^2 +\Omega^2 \mu_\parallel^2 Q^4}
\ ,
\eeq
which demonstrates the scaling form (\ref{eq:corr_rho}) with the scaling exponents from the linear theory (\ref{eq:linexp}), and 
\beq
S^{\rm lin}_{\rho\rho}(y)=
\int  \frac{\dd^d {\bf Q} \dd \Omega}{(2\pi)^{(d+1)}} \frac{2D Q^2 \exp\left[\ii\left(   {\bf Q}\cdot \hat{\br} -\Omega y\right)  \right]}{ (\Omega^2-\zeta Q^4)^2 +\Omega^2 \mu_\parallel^2 Q^4}
 \ .
\eeq
 
As we will show later,  all the scaling exponents from the linear theory 
	(\ref{eq:linexp}) are in fact incorrect  for  describing the hydrodynamic behavior around the MCP due to  the nonlinearities in the EOM.

\subsubsection{ Divergence of correlation length}

Besides the scaling exponents in the correlation functions right at the MCP,  the divergence of the correlation length, as one approaches the MCP, is also governed by 
another set of scaling exponents. For the Ising model, this is the temperature. For the present MCP  however,  this divergence is associated to  two parameters $\alpha_0$, $\kappa_0$. The other two relevant parameters, $\alpha_1$ and $\kappa_1$, take a role akin to the magnetic field in the Ising model.

Since $\kappa_0$ appears in the EOM as a speed of sound for density wave, it is not immediately clear how it might be related to the correlation length. However one can show by rederiving the correlation functions \eqref{eq:lin_response} in the presence of these two couplings, that the equal-time correlation functions are 
\begin{subequations}
\begin{align}
	\label{eq:corr_lin_rho}
    C_\rho(\br,0)&=\int_{ \bq} e^{\ii\bq\cdot \br} \frac{D}{(\alpha_0+\mu_\parallel q^2)(\kappa_0+\zeta q^2)}
     \ , 
    \\
    \bC_g^\bot(\br,0) &=\int_{ \bq} e^{\ii\bq\cdot \br} \frac{D \bP^\bot(\bq)}{\alpha_0+\mu_1 q^2}\ , 
    \\
    \bC_g^\parallel(\br,0)&= \int_{ \bq} e^{\ii\bq\cdot \br} \frac{D \omega^2 \bP^\parallel(\bq)}{\alpha_0+\mu_\parallel q^2} \ .
\end{align}
\end{subequations}
From this standard form it is clear that, in the linear theory, both $\alpha_0^{-1/2}$ and $\kappa_0^{-1/2}$ define a crossover scale, and the larger of the two a correlation length for density correlations, while $\alpha_0^{-1/2}$ is always the correlation length for momentum correlations.

As the divergence of the correlation length is described by these two parameters,  $\alpha_0$ and $\kappa_0$, there are also two exponents  which we call $y_1$ and $y_2$. At the linear level, 
these exponents correspond expectedly to their mean-field values:
\beq
\label{eq:lincorrexp}
y_1^{\rm lin} = y_2^{\rm lin} = 2 
\ .
\eeq
We note that these scaling exponents are again expected to be modified by the nonlinearities, as we shall see in the next section.

\begin{figure}
    \centering
    \includegraphics[width=\linewidth]{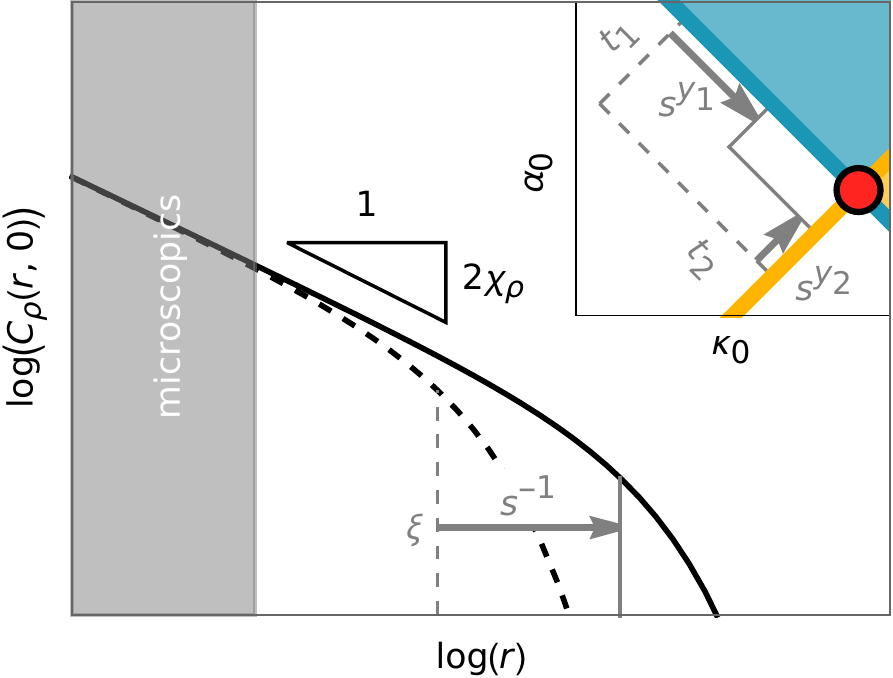}
    \caption{Scaling behavior of the correlation length when approaching the MCP. The inset shows the phase diagram in terms of the couplings $\alpha_0$ and $\kappa_0$ under the assumption that $\alpha_1=\kappa_1=0$, and the main figure shows the equal-time density correlation function $C_\rho(\br,0)$ in log-log scale (black lines) which, if sufficiently close to the MCP (red circle in the inset) shows the critical scaling behavior characterized by the scaling exponent $\chi_\rho$ (slope triangle). In reality though, the system will never be exactly at the critical point. This can be characterized by the distances $t_1$ and $t_2$ (gray lines in inset) from the critical point of disordered phase separation (blue line in inset) and from the critical order-disorder transition (yellow line in inset) respectively. This manifests in a finite correlation length $\xi$ above which the scaling behavior breaks down (gray lines in main figure). As one approaches the fixed point, by decreasing $t_1$ and $t_2$, the correlation length diverges. If this is done carefully, such that the second relation in Eq.~(\ref{eq:corr_length}) remains unchanged, for example by rescaling $t_1$ and $t_2$ by a factor $s^{y_1}$ and $s^{y_2}$ respectively (gray arrow in inset), the correlation length rescales according to Eq.~(\ref{eq:corr_length}), i.e. by a factor $s^{-1}$ (gray arrow in main figure). }
    \label{fig:corr_length}
\end{figure}

Experimentally, these exponents define a relationship between the correlation length $\xi$ and the distances $t_1$ and $t_2$ in the phase diagram (see the inset of Fig.~\ref{fig:corr_length}) from the critical points of disordered phase separation (blue upwards triangle in Fig.~\ref{fig:cartoon} and blue line in the inset of Fig.~\ref{fig:corr_length}) and the critical order-disorder transition (yellow downwards triangle in Fig.~\ref{fig:cartoon} and yellow line in in the inset of Fig.~\ref{fig:corr_length}),
\beq
\label{eq:corr_length}
    \xi \sim t_1^{-\frac{1}{y_1}} \sim t_2^{-\frac{1}{y_2}} \ .
\eeq
This means that, upon approaching the MCP, one has to enforce the relationship given by the right proportionality in Eq.~(\ref{eq:corr_length}) to see a clear scaling behavior in the correlation length.
This relationship is also visualized in Fig.~\ref{fig:corr_length}.

\section{Nonlinear regime}
\label{sec:nonlin}

While the scaling behavior described in Sec.~\ref{sec:linear} is qualitatively expected generally, the critical exponents (\ref{eq:linexp},\ref{eq:lincorrexp}) obtained in the linear theory are only expected to be exact when the spatial dimension $d$ is high enough. 
Below a certain upper critical dimension $d_c$, nonlinear terms become important which modify the scaling and correlation length exponents. The linear exponents (\ref{eq:linexp}) can however be used to gauge the importance of various nonlinearities in the EOM (\ref{eq:TT}) as $d$ is lowered.

We now turn to the full EOM of $\bg$ (\ref{eq:TT}), and perform a rescaling (\ref{eq:rescale}) with the linear exponents (\ref{eq:linexp}). If the spatial dimension $d$ is large enough, all nonlinear terms are irrelevant, i.e., they vanish as $\ell \rightarrow \infty$. As $d$ decreases from, say infinity, the nonlinear terms that first become relevant (and are not fine-tuned to zero), i.e., terms that diverge as $\ell \rightarrow \infty$, are
\beq
\alpha_2 \rho^2 \bg \ \ \ \text{and} \ \ \ \kappa_2 \rho^2 \vnab \rho
\ ,
\eeq
which happens at the upper critical dimension $d_c = 6$. These non-linear terms, together with the linear terms, support the following symmetry:
\beq
\label{eq:symmetry}
\rho \rightarrow -\rho \ \ \ \text{and} \ \ \ \bg \rightarrow -\bg
\ .
\eeq
One can therefore simplify the consideration by restricting to the subspace of EOM compatible with this symmetry, in which $\alpha_1$ and $\kappa_1$ are vanishing and are not generated under RG transformations. Higher-order terms breaking this symmetry are irrelevant close to the upper critical dimension, modifying our results only beyond the order considered in this work. Therefore, close to the upper critical dimension, the symmetry \eqref{eq:symmetry} is a property of the MCP.

A physical interpretation of this emergent symmetry (\ref{eq:symmetry}) corresponds to the equivalence between a high-density band traveling in direction $\hat{\bf n}$ and the corresponding low-density band (with the same profile but inverted) traveling in the $-\hat{\bf n}$ direction.

Just below six dimensions, the universal hydrodynamic EOM (\ref{eq:TT}) is therefore
\beqn
\label{eq:nonlinear}
\nonumber
\gamma \pp_t \bg&=&\mu_1 \nabla^2 \bg +\mu_2\vnab (\vnab \cdot \bg)-\alpha_0 \bg -\kappa_0 \vnab \rho+\bff \\
& & - 
\alpha_2 \rho^2 \bg- \frac{\kappa_2}{3}\vnab \left(\rho^3\right) +\zeta \nabla^2 \vnab \rho
\ ,
\eeqn
where $\gamma$ is a dimensionless coefficient introduced to allow for renormalization of the temporal derivative term associated to $\bg$. Note that such a RG correction to the temporal derivative is also present in a recent study of incompressible active fluids with quenched disorder \cite{chen_a22b}. Further, we 
note that the signs of the nonlinear terms (with $\alpha_2, \kappa_2>0$) are chosen for the sake of stability. By the same token, the term ${\zeta} \nabla^2 \vnab \rho$  stabilizes the system in the case of $\kappa_0<0$. In fact, this term is marginal according to our linear theory and is therefore required in our discussion. It can be interpreted as an effective ``pressure" or compressibility term for the momentum density field which, in the limit of small $\zeta$ manifests as an effective diffusion for the density mode. Its dispersion relation is given as  
\begin{equation}
\label{eq:dispersion}
    \omega_{\rho} =  \frac{ \mu_\parallel}{2\gamma} \left( \ii  -\sqrt{\frac{4\gamma \zeta}{ \mu_\parallel^2}-1} \right)  q^2 \approx \ii \sqrt\frac{\zeta}{\gamma} q^2 \ ,
\end{equation}
where the last approximation is valid in the limit of $\zeta\ll \mu_\parallel^2/(4\gamma) $ such that the diffusion constant of the density mode is given by $\sqrt{\zeta/\gamma}$.

\section{DRG analysis}
\label{sec:drg}

Traditionally, a DRG analysis together with the $\epsilon$-expansion method is now applied. As we will demonstrate now however, a one-loop calculation as is usually performed, will not be sufficient. 

The DRG \cite{forster_pra77} is usually performed by first transforming the EOM (\ref{eq:cont},\ref{eq:nonlinear}) to Fourier-space and then splitting the fields into small and large scale modes, arbitrarily split at an inverse length-scale $\Lambda^\prime=\Lambda e^{-\ell}$, which is a fraction of the physical cutoff scale $\Lambda$ that defines the smallest length scale of the system, e.g., the average distance between individual particles:
\begin{subequations}
\begin{align}
	\rho(\bq,\omega) &= \rho_>(\bq,\omega)+\rho_<(\bq,\omega) \ , \\
	\bg(\bq,\omega) &= \bg_>(\bq,\omega)+\bg_<(\bq,\omega) \ ,
\end{align}
\end{subequations}
such that $\rho_>(\bq,\omega) = \rho(\bq,\omega)$ if $|\bq| > \Lambda^\prime$ and $\rho_>(\bq,\omega) = 0$ otherwise, etc.

One can then eliminate the small scale modes $\rho_>$ and $\bg_>$ from the EOM by recursively reinserting the formal solution for the small scale modes in terms of the large scale modes provided by the EOM. This generates a hierarchy of terms, which can only be truncated by assuming the interaction terms are small, i.e., by going to the perturbative limit. Averaging this expression over the small scale noise terms $f_>$ that remained so far in the equation, then generates effective contributions to the coefficients of the EOM (\ref{eq:nonlinear}) that depend on the coarse-graining scale $\Lambda ^\prime$. These terms can be represented diagrammatically through Feynman diagrams which can be classified by their number of loops, i.e., number of integrals one needs to solve to determine the correction.

The conventions we will be using in this paper are 
\begin{subequations}
\label{eq:feynman_rules}
\begin{align}
\diagram{28} = \bG(\tilde \bq) \ , \\
\diagram{29} = \frac{1}{-\ii \omega_q} \ , \\
\diagram{30} = -\ii \bq \ , \\
\diagram{31} = 2D \ \id \ , \\
\label{eq:valph}
\diagram{32} = \alpha_2\  \id \ , \\
\label{eq:vkap}
\diagram{33} = \ii \bq \kappa_2 \ ,
\end{align}
\end{subequations}
where $\id$ is the identity matrix and $\bG$ is the propagator of the momentum density fields
\begin{subequations}
\label{eq:prop_drg}
\begin{align}
\bG(\tilde{\bq}) &= G_{\parallel}(\tilde{\bq}) \bP^{\parallel}(\bq) + G_{\perp}(\tilde{\bq}) \bP^{\bot}(\bq) \ ,  \\
G_{\parallel}(\tilde{\bq}) &= \frac{-\ii\omega}{-\ii\omega(-\ii { \gamma}\omega+\alpha_0 +\mu_\parallel q^2)+\kappa_0 q^2+\zeta q^4} \ ,
\\
G_{\perp}(\tilde{\bq}) &= \frac{1}{-\ii { \gamma} \omega +\alpha_0+\mu_1 q^2 }\ .
\end{align}
\end{subequations}
In general, each diagram expression is of the tensorial rank equal to the number of open unbroken lines.

These, so-called, graphical corrections are 
\bew
\begin{align}
\label{eq:DRG-graphical-a2}
\id \Gr^{\rm DRG}_{\alpha_2} =& -2\diagram{16}-4\diagram{17}-4\diagram{18}-4\diagram{19}-2\diagram{20}  , \\
\label{eq:DRG-graphical-k2}
\ii\bq\Gr^{\rm DRG}_{\kappa_2} =&  -4\diagram{21}-4\diagram{22}-6\diagram{23}   -4\diagram{24}\\ \nonumber
&-4\diagram{25}-6\diagram{26} , 
\end{align}
\ew
\begin{equation}
\label{eq:DRG-graphical-0}
\id\Gr^{\rm DRG}_{\alpha_0} = \diagram{2}  , \hspace{0.5cm}
\ii\bq\Gr^{\rm DRG}_{\kappa_0} =\diagram{6}  , 
\end{equation}
where outgoing lines without a wave vector imply that the corresponding wavevector has been set to zero. The loop wavevector has not been written explicitly and its integral is implied. Whether the external wavenumber $\bq$ is routed through the upper or the lower part of the loop is irrelevant up to order $\epsilon$. Also, the tildes are omitted since the external frequency $\omega_q$ is set to 0 for every diagram in our approximation. Only internal lines correspond to a propagator according to Eq.~(\ref{eq:feynman_rules}). External lines only mark open vector indices and external wavevectors.  

We further note that the different prefactors in Eq.~(\ref{eq:DRG-graphical-k2}) stem from the fact that the right-hand side has to be expanded to linear order in $\bq$. For diagrams whose first vertex has three outgoing density propagators (dotted lines), i.e. the vertex in Eq.~(\ref{eq:vkap}), it is immaterial at this order whether the external wave vector  leaves before or after the loop part, e.g.,
\begin{equation}
\diagram{37} = \diagram{23} + \mathcal O(q^2).
\end{equation}
For diagrams where the first vertex has two outgoing density fields and one outgoing momentum field (two dotted and one unbroken line), i.e. the vertex in Eq.~(\ref{eq:valph}), contributions where the external wave vector leaves before passing through the loop part vanish identically since they are wave vector independent and the  wave vector independent part vanishes due to antisymmetry of the integrand, e.g.,
\begin{equation}
\diagram{38} = 0.
\end{equation}
This attributes the relative factor $2/3$ between these two types of diagrams since all permutations of the outgoing wavevector need to be considered. 

At the 1-loop level, we can set $\kappa_0$ and $\alpha_0$ to zero on 
the right-hand-side of Eqns.~(\ref{eq:DRG-graphical-a2}) and (\ref{eq:DRG-graphical-k2}). For Eq.~(\ref{eq:DRG-graphical-0}), the expansion must be carried out explicitly to second order in $\kappa_0$ and $\alpha_0$ however, which is most conveniently done after the frequency integration. Further details of the evaluation of the diagrams can be found in App.~\ref{app:drg}.

As an example, for one of the specific 1-loop diagrams from Eq.~(\ref{eq:DRG-graphical-a2}) that provides a RG correction to $\alpha_2$ the analytical expression is,
\beqn
\nonumber
&& \diagram{18}=\alpha_2^2 
    \int_{\Lambda \ee^{ - \ell}}^\Lambda \frac{\dd^d \bp}{(2\pi)^d} \int_{-\infty}^\infty \frac{ \dd \omega_p}{2\pi} \\
\nonumber 
&&\times \frac{2D \ \bp \otimes \bp}{(\gamma\omega_p^2-\zeta p^4)^2+\omega_p^2 \mu_\parallel^2 p^4} \frac{\omega_p  }{\ii(\gamma\omega_p^2-\zeta p^4) +\omega_p \mu_\parallel  p^2}
  \ .
  \eeqn
 We note here that the $\zeta$ term is indeed crucial to regularize the RG calculation, as without it the above 1-loop frequency integral is clearly divergent.

Now together with these corrections, stemming from the nonlinear terms, we can reattempt the rescaling in Eq.~(\ref{eq:rescale}) which we include as an effective rescaling of the couplings. In addition, the EOM (\ref{eq:nonlinear}) is divided by $\gamma$, to fix the time-derivative coefficient to unity:
\begin{subequations}
\begin{align}
\mu_1 &\rightarrow \mu_1  \frac{e^{(z-2) \ell}}{\gamma} \ , \\
\mu_\parallel &\rightarrow \mu_\parallel \frac{e^{(z-2) \ell}}{\gamma} \ , \\
\zeta &\rightarrow \zeta  \frac{e^{(z-3+\chi_\rho-\chi_g) \ell}}{\gamma} \ , \\
D &\rightarrow D  \frac{e^{(- 2\chi_g+z-d) \ell}}{\gamma^2} \ , \\
\alpha_0 &\rightarrow \alpha_0 \frac{e^{z \ell}}{\gamma} \ , \\
\kappa_0 &\rightarrow \kappa_0 \frac{e^{(z+\chi_\rho-\chi_g) \ell}}{\gamma} \ , \\
\alpha_2 &\rightarrow \alpha_2 \frac{e^{(z+2\chi_\rho) \ell}}{\gamma} \ , \\
\kappa_2 &\rightarrow \kappa_2 \frac{e^{(z+3\chi_\rho-\chi_g) \ell}}{\gamma} \ .
\end{align}
\end{subequations}
If we now consider that $\ell$ is an infinitesimal number $\ell\rightarrow \dd \ell$, defining the so-called Wilsonian momentum shell, we can write down the DRG flow equations:
\begin{subequations}
\label{eq:prepre_drg_flow}
\begin{align}
\partial_\ell \mu_1 &= (z-2-\Gr^{\rm DRG}_{\gamma}/\gamma) \mu_1 +\Gr^{\rm DRG}_{\mu_1} \ , \\
\partial_\ell \mu_\parallel &= (z-2-\Gr^{\rm DRG}_{\gamma}/\gamma)\mu_\parallel +\Gr^{\rm DRG}_{\mu_\parallel}  \ , \\
\partial_\ell \zeta &= (z-3+\chi_\rho-\chi_g-\Gr^{\rm DRG}_{\gamma}/\gamma)\zeta +\Gr^{\rm DRG}_{\zeta}\ , \\
\partial_\ell D &= (- 2\chi_g+z-d-2\Gr^{\rm DRG}_{\gamma}/\gamma)D +\Gr^{\rm DRG}_{D} \ , \\
\partial_\ell \alpha_0 &= (z-\Gr^{\rm DRG}_{\gamma}/\gamma) \alpha_0 +\Gr^{\rm DRG}_{\alpha_0} \ , \\
\partial_\ell \kappa_0 &= (z+\chi_\rho-\chi_g-\Gr^{\rm DRG}_{\gamma}/\gamma)\kappa_0 +\Gr^{\rm DRG}_{\kappa_0} \ , \\
\partial_\ell \alpha_2 &= (z+2\chi_\rho-\Gr^{\rm DRG}_{\gamma}/\gamma)\alpha_2+\Gr^{\rm DRG}_{\alpha_2} \ , \\
\partial_\ell \kappa_2 &= (z+3\chi_\rho-\chi_g-\Gr^{\rm DRG}_{\gamma}/\gamma)\kappa_2+\Gr^{\rm DRG}_{\kappa_2} \ .
\end{align}
\end{subequations}
This means that there are three parts that contribute to the flow equation of each coupling: the rescaling of fields, lengths and time, the graphical correction of each term, which have been rescaled by the same factor as the respective coupling, and finally the graphical correction of $\gamma$ from dividing the total EOM by this factor. The coupling $\gamma$ itself is not rescaled
\begin{equation}
\partial_\ell\gamma = \gamma +\Gr^{\rm DRG}_{\gamma}
\end{equation}
and will in fact not approach a fixed point. But this does not matter since the rescaled EOM no longer depend on $\gamma$.

If the flow equations (\ref{eq:prepre_drg_flow}) are vanishing, this means that the EOM are invariant under this rescaling transformation {\it in the presence} of nonlinearities, which implies power- law correlations as discussed in Sec.~\ref{sec:scale_inv}. 

To facilitate the comparison between our DRG calculation and our FRG analysis to be presented later, we further define the dimensionless couplings through which all the flow equations can be expressed,
\begin{subequations}
\label{eq:bar_alpha_drg}
\begin{align}
\bar\mu & = \frac{\mu_1}{\mu_\parallel} \\
\bar\zeta & = \frac{\gamma\zeta}{\mu_\parallel^2} \\
    \bar\alpha_0 &=\frac{\alpha_0}{\mu_\parallel \Lambda^2}\ ,
    \\
    \bar \kappa_0 &= \frac{\kappa_0}{\zeta \Lambda^2} \ ,
    \\
    \bar\alpha_2 &= \frac{\alpha_2 \Lambda^{d-6}DS_d}{\mu_\parallel^2\zeta (2\pi)^d} \ , 
    \\
    \bar{\kappa}_2 &=\frac{\kappa_2 \Lambda^{d-6}DS_d}{\mu_\parallel\zeta^2(2\pi)^d}  \ , 
\end{align}
\end{subequations}
 where the geometric factor $S_d/(2\pi)^d$, with the surface area of a $d$-dimensional unit sphere $S_d=2\pi^{d/2}/\Gamma(d/2)$ and the Euler gamma function $\Gamma$, was introduced for convenience. In this way, the rescaling introduced earlier is removed again from the flow equations, which might seem surprising. However, engineering and scaling dimension are actually very closely related. Any anomalous scaling exponent is generated by the renormalization of couplings that relate different units. For example, the diffusion constant $\mu_\parallel/\gamma$ relates time and length scales. The four couplings carrying that information in this system are $\mu_\parallel$, $\zeta$, $\gamma$ and $D$ which are all removed from appearing explicitly in the flow equations due to  making them  dimensionless (\ref{eq:bar_alpha_drg}). What remains in the flow equations, however, is their graphical corrections
\begin{subequations}
\begin{align}
\eta_\mu &= \frac{\Gr^{\rm DRG}_\mu}{\mu_\parallel} \ , \\
\eta_\rho &= \frac{\Gr^{\rm DRG}_\zeta}{\zeta} +\eta_\gamma-2\eta_\mu \ , \\
\eta_\gamma &= \frac{\Gr^{\rm DRG}_\gamma}{\gamma} \ ,\\
\eta_D &= \frac{\Gr^{\rm DRG}_D}{D} \ .
\end{align}
\end{subequations}

The dimensionless flow equations can then be written as
\begin{subequations}
\label{eq:pre_drg_flow}
\begin{align}
\partial_\ell \bar\mu &= -\eta_\mu \bar\mu +\bar\Gr^{\rm DRG}_{\mu_1}  \ , \\
\partial_\ell \bar\zeta &=  \eta_\rho \bar \zeta \ , \\
\partial_\ell \bar\alpha_0 &= (2-\eta_{\mu}) \bar\alpha_0 +\bar\Gr^{\rm DRG}_{\alpha_0} \ , \\
\partial_\ell \bar\kappa_0 &= (2-2\eta_{\mu}+\eta_\gamma-\eta_\rho)\bar\kappa_0 +\bar\Gr^{\rm DRG}_{\kappa_0} \ , \\
\partial_\ell \bar\alpha_2 &= (6-d-4\eta_\mu+\eta_\gamma+\eta_D-\eta_\rho)\bar\alpha_2+\bar\Gr^{\rm DRG}_{\alpha_2} \ , \\
\partial_\ell \bar\kappa_2 &= (6-d-5\eta_\mu+2\eta_\gamma+\eta_D-2\eta_\rho)\bar\kappa_2+\bar\Gr^{\rm DRG}_{\kappa_2} \ .
\end{align}
\end{subequations}
The dimensionless graphical corrections $\bar \Gr^{\rm DRG}$ are defined analogously to their respective couplings.

Now, since all nonlinearities in the EOM are cubic in nature (the $\alpha_2$ and $\kappa_2$ terms),  the only coefficients that receive graphical corrections are $\alpha_{0,2}$ and $\kappa_{0,2}$ at the 1-loop level, which are shown in Eq.~(\ref{eq:DRG-graphical-a2}-\ref{eq:DRG-graphical-0}). The other graphical corrections are all zero, i.e.,
\begin{equation}
\Gr^{\rm DRG}_{\mu_1}=\Gr^{\rm DRG}_{\mu_\parallel}=\Gr^{\rm DRG}_{\zeta}=\Gr^{\rm DRG}_{D}=\Gr^{\rm DRG}_{\gamma}=0 \ ,
\end{equation}
 and as a result,
\begin{equation}
\eta_\mu=\eta_\rho=\eta_\gamma=\eta_D=0.
\end{equation}
Therefore, we can directly infer that the scaling exponents are unchanged from the linear theory
\begin{equation}
z^{\rm DRG}=2,~~\chi_\rho^{\rm DRG}=\frac{4-d}{2},~~\chi_g^{\rm DRG} =\frac{2-d}{2} \ .
\end{equation}
The remaining flow equations, perturbatively expanded to second order in $\bar\alpha_0$, $\bar\kappa_0$, $\bar\alpha_2$ and $\bar\kappa_2$ are 
\begin{subequations}
\label{eq:drg_flow}
\begin{align}
\partial_\ell \bar\alpha_0 &= 2 \bar\alpha_0 + \bar\alpha_2(1-\bar\alpha_0-\bar\kappa_0) \ , \\
\partial_\ell \bar\kappa_0 &= 2 \bar\kappa_0 + \bar\kappa_2(1-\bar\alpha_0-\bar\kappa_0) \ , \\
\partial_\ell \bar\alpha_2 &= \epsilon\bar\alpha_2 - \bar{\alpha}_2\bar\kappa_2 - \left(\frac{5}{3} \right)\frac{2+3\bar\mu+\bar\mu^2+\bar\zeta}{\bar\mu+\bar\mu^2+\bar\zeta}  \bar\alpha_2^2  ,    \\
\nonumber
\partial_\ell \bar\kappa_2 &=\epsilon \bar\kappa_2-3\bar\kappa_2^2-\frac{11}{3} \bar{\alpha}_2\bar\kappa_2 \\
&\ \ +\left(\frac{2}{3}\right)\frac{5+5\bar\mu+2\bar\mu^2+2\bar\mu^3+2\bar\mu\bar\zeta}{\bar\mu^2+\bar\mu^3+\bar\mu\bar\zeta}\bar\alpha_2^2\ .
\end{align}
\end{subequations}

\floatsetup[figure]{style=plain,subcapbesideposition=top}
\begin{figure*}
\label{fig:drgflow}
\begin{tabular}{ll}
\includegraphics[width=0.49\linewidth,trim=0.5cm 0 0 0]{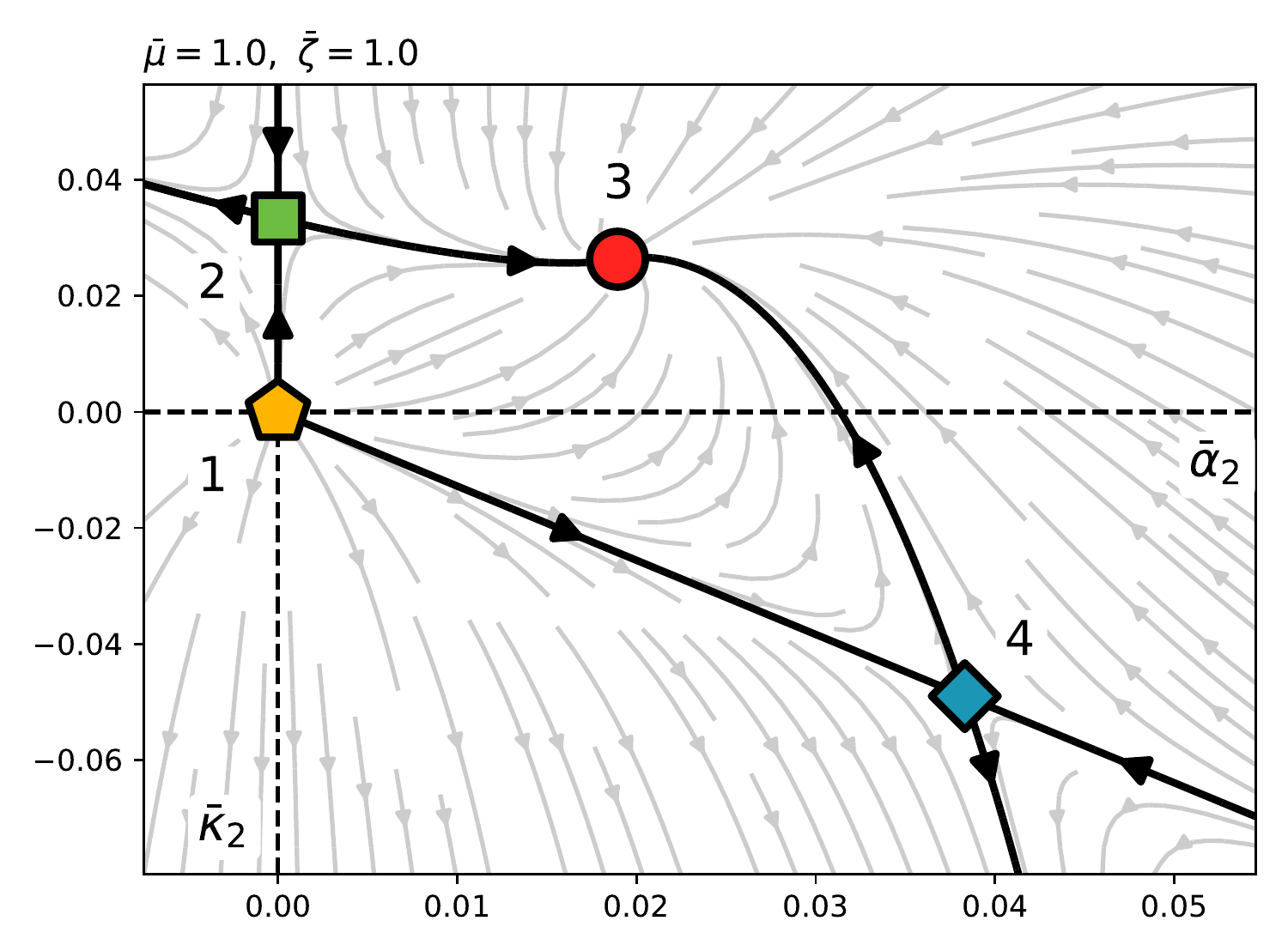} & \includegraphics[width=0.49\linewidth,trim=0.5cm 0 0 0]{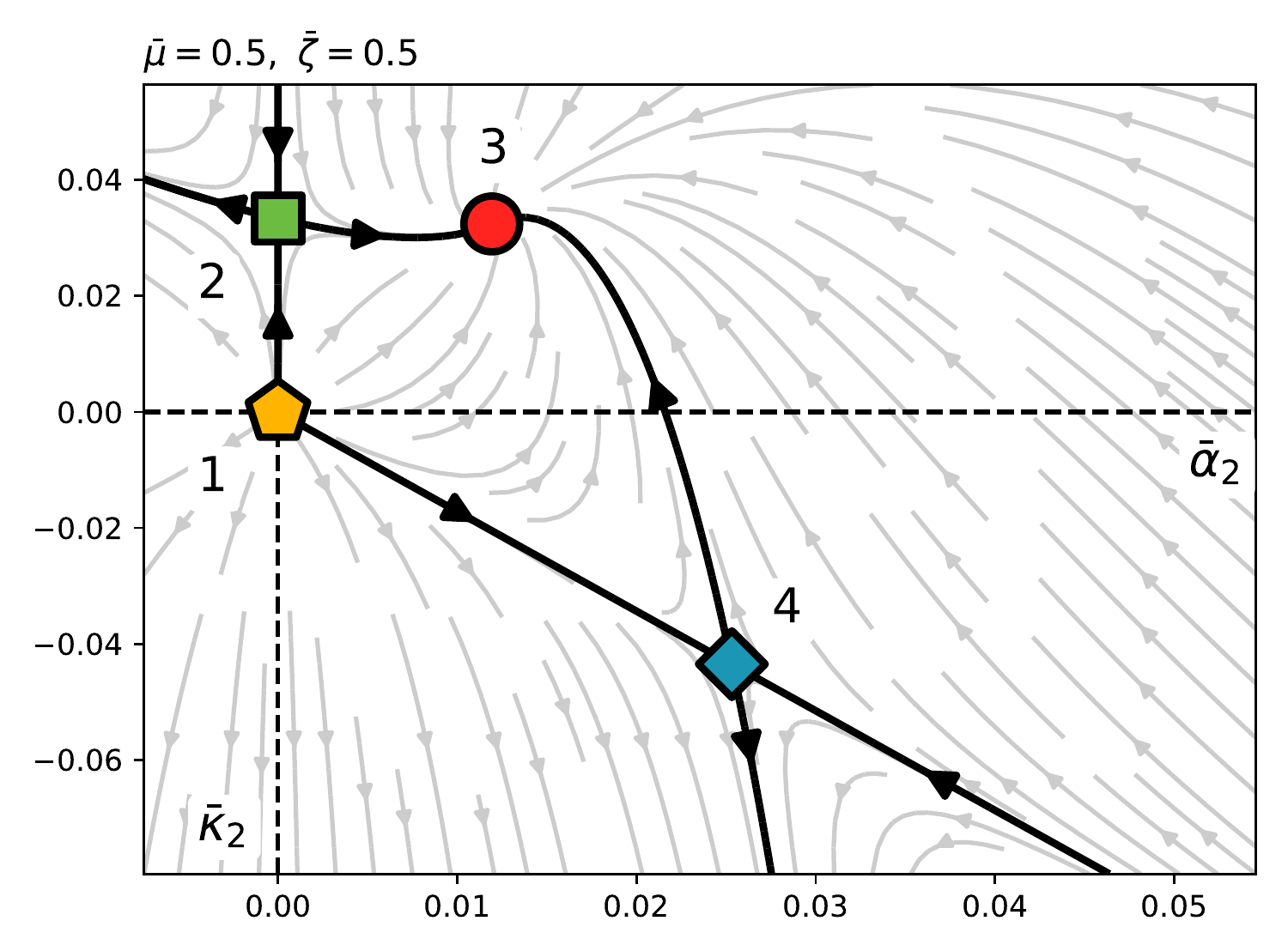}
\end{tabular}

\caption{ It is a priori not clear which values to choose for $\bar \mu$ and $\bar\zeta$ and the flow diagram and the location of the four fixed points, indicated by the yellow pentagon (1), the green square (2), the red circle (3) and the blue diamond (4), depend on this choice. To show this, we plot here two examples of the flow diagram obtained with the different values for $\bar \mu$ and $\bar\zeta$ shown, yielding different results. In both cases we chose $\epsilon=0.1$.}
\end{figure*}

These flow equations have four different fixed point solutions, three of which are nontrivial and depend on the initial values of $\bar\mu$ and $\bar\zeta$. In Fig.~\ref{fig:drgflow}, we show two examples of flow diagrams for different values of $\bar\mu$ and $\bar\zeta$. This suggests that there are potentially three novel universality classes, since for each of them a different rescaling transformation exists, implying different critical exponents.  However, since two of the FP locations, especially that of the attractive one, depend on $\bar\mu$ and $\bar \zeta$, so do the critical exponents.  Since the one-loop DRG calculation predicts no renormalization of these couplings, they can take arbitrary values, according to whichever microscopic model is realized, suggesting that the critical exponents are not universal, which is indicative of the unreliability of this 1-loop calculation.

This  defect can potentially be cured by taking into account two-loop effects, where $\mathcal O(\epsilon^2)$ corrections to the flow equations may lead to nontrivial and universal FP values of the couplings $\bar\mu$ and $\bar \zeta$. As far as we are aware, such a two-loop calculation has never been done explicitly for active matter systems using the DRG formalism.
This may be due to the use of the sharp wavenumber cutoff in the Wilsonian momentum shell regularization that renders a two-loop calculation difficult.
  For similar problems in the literature, one has usually resorted to a field-theoretic approach, where the large scale regularization can be made smooth, see e.g., \cite{frey_pre94}. In this work, we will, however, pursue a functional renormalization group approach instead.

\section{FRG analysis}
\label{sec:frg}

Our FRG analysis is based on the so-called Wetterich equation \cite{wetterich_plb93,morris_ijmpa94,ellwanger_zfpc94}:
\beq
\label{eq:wetterich}
\pp_k \Gamma_k =\frac{1}{2} {\rm Tr} \left[ \left(\Gamma^{(2)}_k +R_k\right)^{-1} \pp_k R_k\right]\ ,
\eeq
where $\Gamma_k$ is the $k$-dependent {\it effective average action}, with $k$ being the inverse length scale up to which fluctuations have been averaged out  (i.e., $k$ plays the equivalent role of $\Lambda' = \Lambda \ee^{- \ell}$ in the previous DRG analysis.) The functional $\Gamma_k$ interpolates from the {\it microscopic} action $\Gamma_{\Lambda}$ to the {\it macroscopic} effective average action $\Gamma_{0}$. As the Legendre transform of the logarithm of the partition function, it contains all statistics of the many-body problem and can therefore be regarded as its full solution. It can further be regarded as the classical action for the average of the fields.  The gradual incorporation of fluctuations as $k \rightarrow 0$ is facilitated by the ``regulator" $R_k$, which serves to suppress fluctuations of length scales greater than $k^{-1}$. The regulator can be chosen arbitrarily as long as $R_\Lambda\approx \infty$ and $R_0=0$ to ensure the correct boundary conditions for $\Gamma_k$. Further, $\Gamma^{(2)}_k$ in Eq.~(\ref{eq:wetterich}) denotes the field dependent matrix of the second functional derivatives of $\Gamma_k$ (i.e., entries are of the form $\delta^2 \Gamma_k/(\delta {\bg} \delta \rho )$, etc), and ${\rm Tr}$ stands for the matrix trace over internal indices and integration over the internal wave vector and frequency.

\subsection{Choice of functional in the Wetterich equation}

While the Wetterich equation (\ref{eq:wetterich}) is in principle {\it exact}, the actual implementation of the RG flow relies on restricting the functional $\Gamma_k$ to a manageable form. Here, we will take $\Gamma_k$ to be the functional obtained from the EOM (\ref{eq:cont},\ref{eq:nonlinear}) via the Martin-Siggia-Rose-de Dominicis-Janssen formalism \cite{martin_pra73,janssen_ZPhB76,dedominics_JPhCol76,canet_jopa11}:
\beqn
\nonumber
\label{eq:G}
&&\Gamma_k[\bar{\bg},\bg, \bar{\rho},\rho] = \int_{\tilde\br} \bigg\{\bar{\rho} \left(\pp_t \rho +\vnab \cdot \bg\right)-D |\bar\bg|^2
\\
\nonumber
&& \ \ 
+\bar{\bg}\cdot \bigg[\gamma \pp_t \bg-\mu_1 \nabla^2 \bg -\mu_2\vnab (\vnab \cdot \bg)
+\alpha_0 \bg 
\\
&&\ \ 
+\kappa_0 \vnab \rho+
\alpha_2\rho^2 \bg+ \kappa_2 \rho^2 \vnab \rho-  \zeta  \nabla^2 \vnab \rho\bigg]\bigg\}\ ,
\eeqn
where $\int_{\tilde\br} \equiv \int \dd^d \br \dd t$, and all coefficients above ($\mu_{1}$, $\mu_{2}$, $\alpha_0$, etc.) are now $k$ dependent.
The response fields introduced by the formalism are denoted by $\bar\bg$ and $\bar\rho$. 

Note the absence of any coefficients in the density `sector' of $\Gamma_k$, i.e., terms proportional to $\bar\rho$ in (\ref{eq:G}). This is due to the fact that it does not renormalize because of an extended symmetry as defined in Ref.~\cite{canet_pre15}. Specifically, this extended symmetry stems from the linear nature of the continuity equation (\ref{eq:cont}), which implies that, under the transformation
\begin{equation}
\label{eq:extsym}
    \bar\rho(\br,t) \rightarrow \bar \rho(\br,t) + \varepsilon(\br,t),
\end{equation}
with an arbitrary field $\varepsilon$,
the microscopic action $\Gamma_\Lambda$ transforms linearly in the fields,
\begin{equation}
    \delta \Gamma_\Lambda = \int_{\tilde \br} \varepsilon ( \partial_t \rho+ \nabla \cdot \bm \bg).
\end{equation}
Since the transformation (\ref{eq:extsym}) is scale-independent it commutes with the scale derivative $\partial_k$ and we can use the Wetterich equation (\ref{eq:wetterich}) to see how this relation changes under RG transformations in the case of infinitesimal $\varepsilon$:
\beqn
\label{eq:rgextsym}
&&\pp_k \delta\Gamma_k =\\
\nonumber
&&-\frac{1}{2} \tr \left[\left(\Gamma^{(2)}_k +R_k\right)^{-1} \delta\Gamma_k^{(2)} \left(\Gamma^{(2)}_k +R_k\right)^{-1} \pp_k R_k\right]\ .
\eeqn
As $\delta \Gamma_k$ is linear initially at the scale $k=\Lambda$, $\delta\Gamma_k^{(2)}$ vanishes, so $\delta \Gamma_k$ remains unchanged at the infinitesimally larger RG scale $k=\Lambda+\dd k$. This argument can be repeated at this scale and so on, showing that $\partial_k\delta \Gamma_k=0$ and $\delta \Gamma_\Lambda = \delta \Gamma_k = \delta\Gamma_0$ at all scales. Thus, the density sector does not renormalize, which is why, in Eq.~(\ref{eq:G}), we have set the coefficients characterizing it to unity, fixing the engineering dimensions of the fields. This nonrenormalization further implies the hyperscaling relationship between the density and momentum density field,
\begin{equation}
	\chi_\rho-\chi_g-1 = z-2 \ .
\end{equation}

For the momentum density `sector' of $\Gamma_k$, i.e., terms proportional to $\bar\bg$ in (\ref{eq:G}), we know from our linear theory that this form of $\Gamma_k$ is sufficient only around the critical dimension $d_c=6$. 
As a result, we expect that the validity of our quantitative predictions is limited to around $d_c$. Therefore, we will express our results as corrections to the linear theory in terms of $\epsilon = d_c-d$.
In particular, our results for universal exponents will coincide with the perturbative DRG results to order $\epsilon$, if the fixed point values for $\bar\mu$ and $\bar\zeta$ are put into the DRG calculation by hand. At the same time, 
corrections of order $\mathcal O(\epsilon^2)$ are expected to differ from the DRG results at the same order, since the FRG analysis is nonperturbative in nature. Even though our approach, therefore,  becomes perturbative in the couplings $\alpha_0$, $\kappa_0$, $\alpha_2$ and $\kappa_2$, since their FP values are controlled by $\epsilon$, our approach is not fully perturbative, since we take the full dependence of the flow equations on $\bar\mu$ and $\bar\zeta$  into account, whose FP values are not controlled by $\epsilon$.

\subsection{Regulator}


Besides the form of the average action, the regulator $R_k$ needs to be specified, which we choose to be, in spatio-temporally Fourier transformed space,
\beq
\label{eq:reg}
\begin{aligned}
 R_k(\tilde{\bq},\tilde{\bp}) &= (2\pi)^{d+1}\delta^{d+1}(\tilde{\bq} + \tilde{\bp}) \\
& \times
	  \begin{pmatrix} 
	    0        & \id A_k(q^2) &0& \ii \bq B_k(q^2) \\
	    \id A_k(q^2) & 0        & 0 & 0 \\
	    0
	    & 0 & 0 & 0\\
	    -\ii \bq B_k(q^2)  & 0 & 0 & 0\\
	    \end{pmatrix}, 
\end{aligned}
\eeq 
where the ordering of the matrix entries is: $(\bar{\bg},\bg,\bar{ \rho}, \rho)$. The choice of a time-independent regulator is common for dynamical systems \cite{canet_jopa11}. Also, this matrix form does not regulate the density sector directly but rather introduces a $k$ dependent ``pressure  term" in the momentum field sector. This regularization sufficiently cuts off large and small scale fluctuations (for appropriate choices of $A_k$ and $B_k$) while following the overall structure of the EOM. In particular, it leaves the extended symmetry unmodified, implying that density remains conserved, even in the regulated theory, which we believe to be crucial to obtain the correct scaling behavior (e.g., the value of the dynamic exponents in dynamic Ising models depends on whether the dynamics are conservative or not \cite{hohenberg_rmp77}). 

We also define the following in Eq.~(\ref{eq:reg}):
\begin{subequations}
\begin{align}
\nonumber
A_k(q^2) &= \mu_{\parallel,k} k^2 \, m(q^2/k^2) \ ,
\\
B_k(q^2) &= \zeta_{k} k^2 \, m(q^2/k^2)\ ,
\\
\label{eq:alg_reg}
m(y) &= a/y \ ,
\end{align}
\end{subequations}
where we write the $k$-dependence of the couplings explicitly and $a$ is an arbitrary positive constant. In principle, all results obtained should be independent of the regulator choice, however, truncating the form of $\Gamma_k$ usually introduces some form of regulator dependence. This dependence can be judged by the $a$-dependence of the critical exponents. It turns out that for an algebraic regulator as in Eq.~(\ref{eq:alg_reg}), the critical exponents are independent of $a$. This is shown numerically below, but an analytical argument has been given in Ref.~\cite{morris_plb94} as well.

We further verify our results using also another class of regulator, a generalization of the Litim regulator \cite{litim_prd01}
\beq
\label{eq:lit_reg}
    m(y) = a (1-y)^4 \Theta(1-y) \ .
\eeq
The fourth order is required to ensure continuous integrands in the RG flow equations, as derivatives of $m(y)$ up to fourth order appear.

\subsection{FRG flow equations}

With the forms of $\Gamma_k$ and $R_k$ defined, one can then use the Wetterich equation (\ref{eq:wetterich}) to project a set of coupled ordinary differential equations (ODEs), one for each coefficient in the functional (\ref{eq:G}). 
For instance, since
\beq
\alpha_{0,k}=
 \frac{1}{VT}\frac{1}{d} {\rm Tr} \left.\frac{\delta^2 \Gamma_k}{\delta \bar{\bg}(\tilde{\bq}) \delta \bg(\tilde{\bq}) } \right|_{\rho=0}\ ,
\eeq
 where $VT=(2\pi)^{d+1}\delta^{d+1}(0)$ is the spatio-temporal volume, we obtain from the Wetterich equation (\ref{eq:wetterich}) that
\bew
\begin{align}
\pp_\ell \alpha_{0,k} = -\alpha_{2,k} \int_{0}^\Lambda \frac{\dd^d \bp}{(2\pi)^d} \int_{-\infty}^\infty \frac{ \dd \omega}{2\pi}
    \frac{4D p^2 \left[([\kappa_0+B_k]p^2+\zeta p^4-\gamma\omega^2)p^2\partial_\ell B_k(p^2)+(\mu_\parallel p^2+\alpha_0+A_k(p^2))\omega^2\partial_\ell A_k(p^2)\right] }{\left[([\kappa_0+B_k(p^2)] p^2+ \zeta p^4-\gamma\omega^2)^2+\omega^2 (\mu_\parallel p^2+\alpha_0+A_k(p^2))^2\right]^2}
\end{align}
\ew
where $\ell \equiv - \ln (k/\Lambda)$.
The full set of such FRG flow equations correspond to the graphical corrections to the RG flow equations in our previous DRG analysis (\ref{eq:DRG-graphical-a2}-\ref{eq:DRG-graphical-0}). 
At the same time, as in our DRG analysis (\ref{eq:bar_alpha_drg}), it is convenient to introduce dimensionless couplings, for the determination of potential RG fixed points (FPs). In the FRG formalism, the non-dimensionalization and rescaling are performed in a single step, where the inverse scale $k$ takes the role of $ e^{-\ell} \Lambda$. This essentially skips the step where the flow equations are written as in Eq.~(\ref{eq:prepre_drg_flow}). Specifically, we define the following:
\allowdisplaybreaks
\begin{subequations}
\label{eq:bar_alpha_frg}
\begin{align}
\bar\mu & = \frac{\mu_1}{\mu_\parallel} \ , \\ 
\bar\zeta & = \frac{\gamma\zeta}{\mu_\parallel^2} \ , \\ 
    \bar\alpha_0 &=\frac{\alpha_0}{\mu_\parallel k^2}\ ,
    \\ 
    \bar \kappa_0 &= \frac{\kappa_0}{\zeta k^2} \ ,
    \\ 
    \bar\alpha_2 &= \frac{\alpha_2 k^{d-6}DS_d}{\mu_\parallel^2\zeta (2\pi)^d} \ , 
    \\
    \bar{\kappa}_2 &=\frac{\kappa_2 k^{d-6}DS_d}{\mu_\parallel\zeta^2(2\pi)^d}  \ .
\end{align}
\end{subequations}
The rescaling is mostly prescribed by the dimensionality of the couplings. However, we have taken the liberty to rescale $\bar\kappa_0$ and $\bar \kappa_2$ with a factor that contains the dimensionless coupling $\bar \zeta$. This particular choice ensures that the flow equations remain regular in the case that $\bar\zeta\rightarrow 0$.
\allowdisplaybreaks[0]

Like in the DRG calculation, we can now again define the anomalous scaling dimensions
\begin{subequations}
\label{eq:defanomdim}
\begin{align}
\partial_\ell \mu_\parallel &= \eta_\mu \mu_\parallel = \Gr^{\rm FRG}_{\mu_\parallel} \ , \\
\partial_\ell D &=\eta_D D  =  \Gr^{\rm FRG}_{D} \ , \\
\partial_\ell \gamma &=\eta_\gamma \gamma =  \Gr^{\rm FRG}_{\gamma}  \ , 
\end{align}
\end{subequations}
which enter the flow equations of the other couplings through the nondimensioning.
The FRG flow equations can thus be written in a similar fashion as in Eq.~(\ref{eq:pre_drg_flow}):
\begin{subequations}
\label{eq:pre_frg_flow}
\begin{align}
\partial_\ell \bar\mu &= -\eta_\mu \bar\mu +\bar\Gr^{\rm FRG}_{\mu_1}  \ , \\
\partial_\ell \bar\zeta &= (\eta_\gamma-2\eta_\mu+\bar\Gr^{\rm FRG}_{\zeta})\bar\zeta = \eta_\rho \bar \zeta \ , \\
\partial_\ell \bar\alpha_0 &= (2-\eta_{\mu}) \bar\alpha_0 +\bar\Gr^{\rm FRG}_{\alpha_0} \ , \\
\partial_\ell \bar\kappa_0 &= (2-2\eta_{\mu}+\eta_\gamma-\eta_\rho)\bar\kappa_0 +\bar\Gr^{\rm FRG}_{\kappa_0} \ , \\
\partial_\ell \bar\alpha_2 &= (6-d-4\eta_\mu+\eta_\gamma+\eta_D-\eta_\rho)\bar\alpha_2+\bar\Gr^{\rm FRG}_{\alpha_2} \ , \\
\partial_\ell \bar\kappa_2 &= (6-d-5\eta_\mu+2\eta_\gamma+\eta_D-2\eta_\rho)\bar\kappa_2+\bar\Gr^{\rm FRG}_{\kappa_2} \ .
\end{align}
\end{subequations}
Note that the flow equation of $\bar\zeta$ admits two different kinds of FPs: either $\eta_\rho=0$ and $\bar\zeta\neq0$ or  $\eta_\rho\neq0$ and $\bar\zeta=0$. This motivates our choice of dimensionless couplings.

The nondimensional graphical corrections to the FRG flow equations can be written in terms of the second-order functional derivatives:
\begin{subequations}
\label{eq:fs}
\begin{align}
    &\bar\bF_g = \frac{1}{VT}\frac{-k}{\mu_\parallel k^2} \left. \frac{\delta^2 \partial_k \Gamma_k}{\delta \bar{\bg}(\tilde{\bq}) \delta \bg(-\tilde{\bq}) } \right|_{\rho=\rho_{\rm unif}} \ ,\\
    &\bar\bF_\rho = \frac{1}{VT}\frac{-k}{\zeta k^2} \left.\frac{\delta^2 \partial_k \Gamma_k}{\delta \bar{\bg}(\tilde{\bq}) \delta \rho(-\tilde{\bq}) } \right|_{\rho=\rho_{\rm unif}} \ ,\\
    &\bar\bF_D =\frac{1}{VT} \frac{-k}{D} \left.\frac{\delta^2 \partial_k \Gamma_k}{\delta \bar{\bg}(0) \delta \bar{\bg}(0) } \right|_{\rho=\rho_{\rm unif}} \ ,
\end{align}
\end{subequations}
 evaluated at vanishing fields $\bar\bg=\bg=\bar\rho=0$, except for the density which is set to a value $\rho_{\rm unif}$, uniform in space and time, or their dimensionful equivalents, $\bF_g = \mu_\parallel k^2 \bar\bF_g$, $\bF_\rho =  \zeta k^2 \bar\bF_\rho$ and $\bF_D= D\bar\bF_D$. 
Akin to the DRG, the $\bF$ terms can be represented diagrammatically, hence we call them graphical corrections too. Together with the detailed analytical expressions, they are derived in App.~\ref{app:floweq}. 

Before we proceed to detail what these graphical corrections are exactly, we will first discuss how FRG enables us to go beyond the 1-loop calculation in the previous DRG calculation. The strategy that we use here follows from Ref.~\cite{delamotte_lnip12}, which is to evaluate the above $\bar\bF$'s  at a non-vanishing (i.e., off-critical) density.
This procedure is supported by the following physical argument: The effective average action at a nonzero scale $k$, $\Gamma_k$, serves as an effective theory that describes subsystems of size $k^{-1}$. Within this subsystem, the mean density background can be different from the total density background  which is vanishing at the MCP, $\la\rho\ra_k\neq\la\rho\ra_0 =0$, as mass can be exchanged between the subsystems. This is the case when a linear stability analysis on $\Gamma_k$, which includes nonlinear effects on scales smaller than $k^{-1}$, predicts that the homogeneous state $\rho=0$ is unstable, which happens when $\alpha_{0,k}<0$ and/or $\kappa_{0,k}<0$ (the scale dependence has been made explicit here to emphasize that the instability is scale dependent). Then, induced by fluctuations, the system will locally phase separate and spontaneously select a new local density until local stability is reached again. The new local equilibrium is reached  when $\alpha(\rho)=\alpha_0+\alpha_2\rho^2=0$ and $\kappa(\rho)=\kappa_0+\kappa_2\rho^2>0$ or $\kappa(\rho)=0$ and $\alpha(\rho)>0$, whichever happens first. Since this is the physical state of the system, we choose it as the constant background field value for $\rho_{\rm unif}$ when evaluating the second-order  functional derivatives at scale $k$ (\ref{eq:fs}) for the flow equations for $\mu_1$, $\mu_\parallel$ and $\zeta$, i.e.,
\beq
\label{eq:rhounif}
\rho_{\rm unif} = \left\{ \begin{array}{ccc}
     {\rm max}\left( \sqrt{\left|\frac{\alpha_0}{\alpha_2}\right|},\sqrt{\left|\frac{\kappa_0}{\kappa_2}\right|} \right) &\text{if}\ \frac{\alpha_0}{\alpha_2}<0\ \text{and}\ \frac{\kappa_0}{\kappa_2}<0  \\
     \sqrt{\left|\frac{\alpha_0}{\alpha_2}\right|} &\text{if}\ \frac{\alpha_0}{\alpha_2}<0\ \text{and}\ \frac{\kappa_0}{\kappa_2}>0 \\
     \sqrt{\left|\frac{\kappa_0}{\kappa_2}\right|} &\text{if}\ \frac{\alpha_0}{\alpha_2}>0\ \text{and}\ \frac{\kappa_0}{\kappa_2}<0  \\
     0 &\text{otherwise\ .}
\end{array}\right.
\eeq
%
For this definition to work, the flow equations for $\alpha_0$, $\kappa_0$, $\alpha_2$ and $\kappa_2$ must be evaluated at $\rho_{\rm unif}=0$ though, e.g.,
\beq
\alpha_0= \left[ \tr\;\frac{\delta^2 \Gamma_k }{\delta \bar{\bg}(\tilde{\bq}) \delta \bg(-\tilde{\bq})} \right]_{{\rho=0,}\tilde \bq=0}
 \ ,
\eeq
as 
\beq
\left[ \tr\;\frac{\delta^2 \Gamma_k }{\delta \bar{\bg}(\tilde{\bq}) \delta \bg(-\tilde{\bq})} \right]_{{ \rho=\rho_{\rm unif},}\tilde \bq=0}
= 0 \ ,
\eeq
by definition. The procedure seemingly reintroduces the interaction terms that would also be introduced by the couplings $\alpha_1$ and $\kappa_1$, which we excluded above since we were restricting ourselves to the theory obeying the symmetry \eqref{eq:symmetry}. This is however not the case and is instead solely an effect of the projection. The interaction terms are not free variables of the RG-flow but are in fact always fixed by the relationship \eqref{eq:rhounif}. The same effect takes place in Ref.~\cite{delamotte_lnip12} where new effective interaction terms appear that seemingly break the Ising or $O(N)$ symmetry and ultimately yield a nontrivial anomalous dimension.

Note also that distinct from the procedure applied to the Ising model in Ref.~\cite{delamotte_lnip12}, we do not make the change of variables: $\alpha_0\rightarrow\rho_{\rm unif}$, in the flow equations. Instead, we treat $\rho_{\rm unif}$ as an auxiliary variable that is determined from Eq.~(\ref{eq:rhounif}) at each RG step. Our procedure is more advantageous here because of the following: since $\alpha_0$, $\kappa_0$, $\alpha_2$ and $\kappa_2$ are always changing smoothly in $k$, the flow equations never become singular, whereas, since $\rho_{\rm unif}$ can have cusps, e.g., if $\alpha_0$ flips its sign while $\kappa_0>0$, the flow equation for $\rho_{\rm unif}$ has singular behavior at $\rho_{\rm unif} = 0$.

Finally, we can then write the graphical corrections to Eq.~(\ref{eq:pre_frg_flow}) as 
\begin{subequations}
\label{eq:frg_graphical}
\begin{align}
    \label{eq:flowa0}
    \bar\Gr^{\rm FRG}_{\alpha_0} &= \frac{1}{d} \left[ \tr\;\bar\bF_g \right]_{\tilde \bq=0}^{\rho_{\rm unif}=0} \ ,\\
\bar\Gr^{\rm FRG}_{\kappa_0} &= \left[\frac{\bar\bq}{\ii \bar q^2} \cdot \bar\bF_\rho \right]_{\tilde \bq=0}^{\rho_{\rm unif}=0} \ ,\\
    \bar\Gr^{\rm FRG}_{\alpha_2} &= \frac{1}{2d} \frac{\partial^2}{\partial \bar\rho_{\rm unif}^2}\left[ \tr \; \bar\bF_g \right]_{\tilde \bq=0}^{\rho_{\rm unif}=0} \ ,\\
    \bar\Gr^{\rm FRG}_{\kappa_2} &= \frac{1}{2}\frac{\partial^2}{\partial \bar\rho_{\rm unif}^2}\left[ \frac{\bar\bq}{\ii \bar q^2} \cdot \bar\bF_\rho \right]_{\tilde \bq=0}^{\rho_{\rm unif}=0} \ ,\\
    \bar\Gr^{\rm FRG}_{ \mu_1} &= \frac{1}{d-1} \frac{1}{2}\frac{\partial^2}{\partial \bar q^2}\left[ \tr {\bP}^\perp(\bar \bq)\bar\bF_g \right]_{\tilde \bq=0} \ ,\\
    \nonumber
    \bar\Gr^{\rm FRG}_{\zeta} &=\frac{1}{2}\frac{\partial^2}{\partial \bar q^2}\left[ \frac{\bar \bq}{\ii \bar q^2} \cdot \bar\bF_\rho \right]_{\tilde \bq=0} \ , \\
    \bar\Gr^{\rm FRG}_{\mu_\parallel} &= \frac{1}{2}\frac{\partial^2}{\partial \bar q^2}\left[ \frac{1}{\bar q^2} \bar\bq \cdot\bar\bF_g \cdot \bar\bq \right]_{\tilde \bq=0} \ ,\\
    \label{eq:flowgam}
    \bar\Gr^{\rm FRG}_{\gamma} &= \frac{\ii}{d} \frac{\partial}{\partial \bar\omega}\left[ \tr \; \bar\bF_g \right]_{\tilde \bq=0} \ , \\
    \label{eq:flowD}
    \bar\Gr^{\rm FRG}_{D} &= -\frac{1}{2d} \left[ \tr \; \bar\bF_D \right] \ ,
\end{align}
\end{subequations}
where the
$\bar\bF$'s are given in Eq.~(\ref{eq:fs}), and
we have introduced the dimensionless wavevector $\bar \bq = \bq/k$, $\bar q=|\bar\bq|$, and frequency $\bar\omega =\omega{\mu_\parallel}/(\zeta k^2 )$. As described above, the couplings $\alpha_0$, $\kappa_0$, $\alpha_2$ and $\kappa_2$ are evaluated at vanishing background density fluctuation, while the remaining couplings and anomalous dimensions are evaluated at the value shown in Eq.~(\ref{eq:rhounif}). 

We now impose the small $\epsilon$ expansion by neglecting all but the terms from leading order in $\epsilon$ by using
\begin{subequations}
\begin{align}
\bar\mu_\parallel &\sim \epsilon^0 \ , \ \ \ 
&\bar\zeta &\sim \epsilon^0 \ , \\
\bar\alpha_0 &\sim \epsilon \ , \ \ \ 
&\bar\kappa_0 &\sim \epsilon \ ,\\
\bar\alpha_2 &\sim \epsilon \ , \ \ \ 
&\bar\kappa_2 &\sim \epsilon \ ,\\
\eta_\gamma &\sim \epsilon^2 \ , \ \ \ 
&\eta_\mu &\sim \epsilon^2 \ ,\\
\eta_D &\sim \epsilon^2 \ , \ \ \ 
&\eta_\rho &\sim \epsilon^2 \ ,
\end{align}
\end{subequations}
which we know already from our DRG analysis. The flow equations, akin to Eq.~(\ref{eq:drg_flow}) in the DRG analysis, can then be written as 
\begin{subequations}
\label{eq:frg_flow}
\begin{align}
    \partial_\ell \bar\alpha_0 &= 2\bar\alpha_0 + \frac{1}{d} \left[ \tr\;\bar\bF_g \right]_{\tilde \bq=0}^{\rho_{\rm unif}=0} +\mathcal O(\epsilon^2) \ ,\\
\partial_\ell \bar\kappa_0 &= 2 \bar\kappa_0 + \left[\frac{\bar\bq}{\ii \bar q^2} \cdot \bar\bF_\rho \right]_{\tilde \bq=0}^{\rho_{\rm unif}=0} +\mathcal O(\epsilon^2)\ ,\\
    \partial_\ell \bar\alpha_2 &= \epsilon \bar\alpha_2 + \frac{1}{2d} \frac{\partial^2}{\partial \bar\rho_{\rm unif}^2}\left[ \tr \; \bar\bF_g \right]_{\tilde \bq=0}^{\rho_{\rm unif}=0} +\mathcal O(\epsilon^3)\ ,\\
    \partial_\ell \bar\kappa_2 &= \epsilon \bar\kappa_2 + \frac{1}{2}\frac{\partial^2}{\partial \bar\rho_{\rm unif}^2}\left[ \frac{\bar\bq}{\ii \bar q^2} \cdot \bar\bF_\rho \right]_{\tilde \bq=0}^{\rho_{\rm unif}=0} +\mathcal O(\epsilon^3)\ ,\\
    \partial_\ell \bar \mu &= -\eta_\mu \bar\mu + \frac{1}{d-1} \frac{1}{2}\frac{\partial^2}{\partial \bar q^2}\left[ \tr {\bP}^\perp(\bar \bq)\bar\bF_g \right]_{\tilde \bq=0} +\mathcal O(\epsilon^3)\ ,\\
    \nonumber
    \partial_\ell \bar\zeta &= \eta_\rho \bar \zeta \\
    & =(\eta_\gamma-2\eta_\mu)\bar\zeta + \frac{1}{2}\frac{\partial^2}{\partial \bar q^2}\left[ \frac{\bar \bq}{\ii \bar q^2} \cdot \bar\bF_\rho \right]_{\tilde \bq=0} + \mathcal O(\epsilon^3)\ ,
\end{align}
\end{subequations}
where the anomalous dimensions are
\begin{subequations}
\label{eq:anom_dim_flow}
\begin{align}
    \eta_\mu &= \frac{1}{2}\frac{\partial^2}{\partial \bar q^2}\left[ \frac{1}{\bar q^2} \bar\bq \cdot\bar\bF_g \cdot \bar\bq \right]_{\tilde \bq=0} +\mathcal O(\epsilon^3)\ ,\\
    \label{eq:flowgam}
    \eta_\gamma &= \frac{\ii}{d} \frac{\partial}{\partial \bar\omega}\left[ \tr \; \bar\bF_g \right]_{\tilde \bq=0} +\mathcal O(\epsilon^3)\ , \\
    \label{eq:flowD}
    \eta_D &= -\frac{1}{2d} \left[ \tr \; \bar\bF_D \right]_{\tilde \bq=0} +\mathcal O(\epsilon^3)\ .
\end{align}
\end{subequations}

The evaluation of the $\bar\bF$'s at a nonvanishing density can now be seen to serve two purposes: First, it enables the flow equations for the nonlinear couplings $\bar\alpha_2$ and $\bar\kappa_2$ to be projected from $\bar \bF_g$ and $\bar \bF_\rho$ by taking a second-order derivative with respect to $\rho_{\rm unif}$. Secondly, if we were to set $\rho=0$ for the evaluation of the flow equations of $\bar\mu$ and $\bar\zeta$ as well as the anomalous dimensions, they would be vanishing. Then the flow equations would be equivalent to the one-loop DRG result. (In fact, we will show in App.~\ref{app:drg} that one can obtain the one-loop DRG equations exactly from the FRG formalism by using a specific ``sharp" regulator, as in Ref.~\cite{morris_NPhB96}.) Choosing a nonvanishing $\rho_{\rm unif}$, therefore, allows us to incorporate effects that go beyond the one-loop level.

\section{Novel RG fixed points}
\label{sec:fps}

The fixed points of the FRG flow equations (\ref{eq:frg_flow},\ref{eq:anom_dim_flow}) determine the universality classes of the system and their associated scaling behavior. While the flow  equations (\ref{eq:frg_flow},\ref{eq:anom_dim_flow}) can in principle be expressed analytically, the number of terms involved renders them unilluminating. Further, we are unable to analytically solve some of the integrals buried in the definitions of the $\bF$'s (\ref{eq:fsdetailed}). We, therefore, use a combination of computer algebra and numerical methods to solve the FRG equations and thus 
discern the flow of these couplings upon decreasing the inverse length scale $k$. Details of the implementation are given in App.~{\ref{app:numerics}.

In a typical perturbative DRG calculation to one-loop order, one would find that the flow equations for the non-linear couplings $\kappa_2$ and $\alpha_2$ decouple from the relevant couplings $\kappa_0$ and $\alpha_0$. Their FP values can therefore be easily obtained even in a numerical calculation since usually at least one FP in this subspace is attractive. 
In our FRG approach, however, the flow equations for the amplitude ratios $\bar\mu$ and $\bar\zeta$, which the non-linear couplings depend on, are directly proportional to the relevant couplings $\bar\alpha_0$ and $\bar\kappa_0$ through Eq.~(\ref{eq:rhounif}). Therefore, one has to solve all flow equations simultaneously. This is problematic since the relevant couplings diverge from the FP. To tackle this problem in an FRG calculation, one typically invokes the shooting method \cite{berges_pr02,delamotte_lnip12} to fine-tune the relevant parameters, which however becomes difficult when there are many parameters to fine-tune. Here we have developed the following simple method to tackle this problem.

\subsection{Fine-tuning by reversing RG flows}

To steer the couplings towards the fixed points, we \it invert} the sign of the relevant flow equations. This operation manifestly leaves the locations of the FPs invariant, but changes their stability. 
The flow equations, therefore, fine-tune themselves. Once the fixed point solution is found, the original signs can be restored to obtain the critical exponents. This method can also be extended to explore other unstable FPs by inverting additional flow equations.

\setlength{\tabcolsep}{7pt}
\renewcommand{\arraystretch}{1.3}	
	\begin{table*}
	 \caption{Fixed point values for all four fixed points, expressed as an $\epsilon$-expansion from the upper dimension $d_c=6$. The fixed point values of $\alpha_0$ and $\kappa_0$ are normalized such that they are independent of the regulator parameter $a$. The universal amplitude ratios $\bar \mu$ and $\bar\eta$ as well as the anomalous dimensions, the $\eta$'s, are universal.  When no value for $\bar\mu$ or $\bar \zeta$ is given, they can take any arbitrary value and are not universal in this case.
	 }
		\label{tab:fps}
        \begin{tabular}{ccccc || cc || cccc}
        \hline
        \hline
        FP & $\bar\alpha_0/a^{1/2}$ & $\bar\kappa_0/a^{1/2}$ & $\bar\alpha_2$ & $\bar\kappa_2$ &  $\bar\mu$ & $\bar \zeta$ & $\eta_\mu$     & $\eta_\gamma$ & $\eta_D$ & $\eta_\rho$   \\
        \hline
        1 &  $0$&  $0$&  $0$&  $0$ &  &  & $0$               &   $0$               & $0$               & $0$    \\
        2 &  $0$& $-0.399\epsilon$ & $0$ & $0.333\epsilon$ & & $1.43$ & $0$               &   $0$               & $0$               & $0$\\
        3 & ${ -0.277\epsilon}$& ${ -0.285\epsilon }$ &  ${ 0.230\epsilon}$ & ${ 0.237\epsilon}$ & $1.45$ & $1.37$ & $0.033\epsilon^2$ &   $0.044\epsilon^2$ & ${ 0.121\epsilon^2}$ & ${ 0}$ \\
        4 & $-0.118\epsilon$& ${ 0.307\epsilon}$ & $0.101\epsilon$& ${ -0.263\epsilon}$ & $0.31$ & $0$& $0.065\epsilon^2$ &   ${ 0.149\epsilon^2}$ & ${ 0.187\epsilon^2}$ & $-0.083\epsilon^2$  \\
        \hline
        \hline
        \end{tabular}
    \end{table*}

\begin{figure}
		\begin{center}
			\includegraphics[width=\columnwidth]{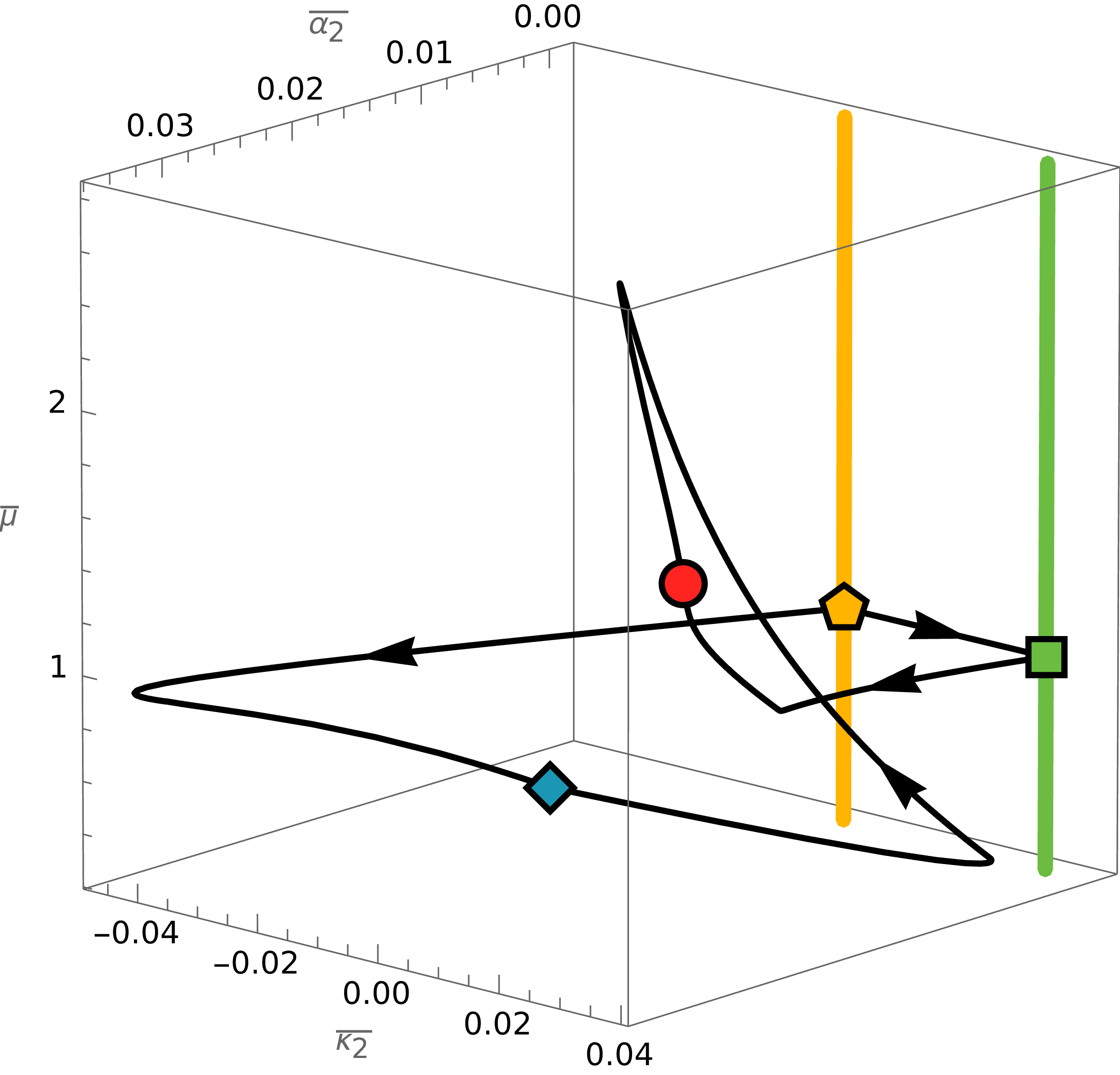}
		\end{center}
		\caption{A projection of the RG flow diagram on the ``critical manifold" to the space spanned by $\bar \alpha_2$, $\bar \kappa_2$ and $\bar \mu$. Our FRG analysis enables us to find four fixed points (FPs): one  is stable (FP3, denoted by the red circle) and three are unstable (FP1, 2, and 4, denoted by the yellow pentagon, green square, and blue diamond, respectively). In this projection, FP1 and FP2 constitute lines of fixed points (yellow and green, respectively). The marked yellow pentagon and green square show the specific FPs reached corresponding to our choice of initial conditions ($\bar\mu=1$ and $\bar \zeta=1$, see App.~\ref{app:numerics}).
		}
		\label{fig:RGflow}
	\end{figure}

With the help of this simple trick, we find a total of four FPs (Fig.~\ref{fig:RGflow} and Tab.~\ref{tab:fps}). One, FP3, is stable and therefore governs generically the universal critical behavior of the MCP under consideration. It is denoted by the red circle in Fig.~\ref{fig:RGflow} and reached by performing the following inversions
\beq
\label{eq:inversion}
    \partial_\ell \bar \alpha_0 \rightarrow -\partial_\ell \bar \alpha_0 \ , \hspace{1cm} \partial_\ell \bar \kappa_0 \rightarrow -\partial_\ell \bar \kappa_0 \ .
\eeq
Here, ``stability" refers to the stability within the ``critical manifold".

We also obtain two other unstable nontrivial FPs: FP2, the green square in Fig.~\ref{fig:RGflow} reached by, additionally to (\ref{eq:inversion}), inverting
\beq
    \partial_\ell \bar \alpha_2 \rightarrow -\partial_\ell \bar \alpha_2 \ ,
\eeq
and FP4, the blue diamond in Fig.~\ref{fig:RGflow} reached by, additionally to (\ref{eq:inversion}), inverting
\beq
    \partial_\ell \bar \kappa_2 \rightarrow -\partial_\ell \bar \kappa_2 \ .
\eeq
Finally, there is the trivial Gaussian FP, FP1, denoted by the yellow pentagon.
 
To the best of our knowledge, the universality classes associated to all FPs are novel, except for the Gaussian FP (yellow pentagon).

\subsection{Genuine nonequilibrium UCs}

In equilibrium the fluctuation dissipation theorem implies $\eta_\gamma=\eta_D$. Since this is clearly broken for FP3 and FP4 (see Tab.~\ref{tab:fps}), we can conclude that FP3 and FP4 {\it are} novel nonequilibrium universality classes.
While the fluctuation dissipation theorem does not seem to be broken for FP2, this does not necessarily imply that FP2 describes the critical phenomenon of an equilibrium system. We discuss this further in Sec.~\ref{sec:fp2}.

\subsection{Nonlinear scaling}
\label{sec:nlscaling}

Now that we have actually found fixed points through our RG analysis, we will revisit the scaling behavior of the theory as well as the correlation functions.

At  the FPs, the anomalous dimensions, i.e., the $\eta$'s (\ref{eq:defanomdim}), take on universal FP values. Therefore, under a RG transformation from a reference scale $k^\prime$, where the system is already sufficiently close to the FP, to the scale $k$, the dimensionful EOM (\ref{eq:nonlinear}) in our truncation transform as
\beqn
\label{eq:nonlinscalingeom}
\nonumber
\gamma^\prime e^{\eta_\gamma \ell} \pp_t \bg&=&\mu_\parallel^\prime e^{\eta_\mu \ell}\left[ \bar\mu^* \nabla^2 \bg +(1-\bar\mu^*) \vnab (\vnab \cdot \bg) \right. \\
\nonumber
&&\left.-\bar\alpha_0^*e^{-2 \ell} \bg- \alpha_2^*e^{(d-6+3\eta_\mu-\eta_D-\eta_\gamma+\eta_\rho)\ell} \rho^2 \bg \right]
 \\
&&+\zeta^\prime e^{(2\eta_\mu+\eta_\rho-\eta_\gamma)\ell}\left[\nabla^2 \vnab \rho-\kappa_0^*e^{-2 \ell} \vnab \rho\right.
\nonumber
\\
&&\left.- \kappa_2^* e^{(d-6+3\eta_\mu-\eta_D-\eta_\gamma+\eta_\rho)\ell} \rho^2 \vnab\rho\right]+e^{\frac{\eta_D\ell}{2}}\bff
\nonumber
\ , \\
\eeqn
where $\ell=\log(k^\prime/k)$, primed couplings denote couplings at the reference scale $k^\prime$ and starred couplings denote the FP value of the couplings. This EOM together with the continuity equation (\ref{eq:cont}) is scale-invariant if we rescale lengths, time and fields,
\beq
\label{eq:rescale_2}
\br \to \br\ee^{\ell},~~t\to t \ee^{z\ell},~~ \rho \to \rho \ee^{\chi_\rho \ell},~~\bg \to \bg \ee^{\chi_g \ell}
\,,
\eeq
with the nonlinear scaling exponents:
\begin{subequations}
\label{eq:nlexponents}
\begin{align}
\label{eq:dynz}
z &= 2-\eta_\mu+\eta_\gamma\ , \\
\label{eq:chig}
\chi_g &= \frac{2-d-\eta_\mu-\eta_\gamma+\eta_D}{2}\ , \\
\label{eq:chirho}
\chi_\rho &= \frac{4-d-3\eta_\mu+\eta_\gamma+\eta_D- \eta_\rho}{2} \ .
\end{align}
\end{subequations}
Note that this only works if $\eta_\rho=0$, which,  for the nontrivial FPs, is  only the case  for  FP2 and FP3, but not FP4. We will discuss the case $\eta_\rho\neq 0$ for FP4 in Sec.~\ref{sec:two-timescales}.

If $\eta_\rho=0$, we have found a rescaling transformation under which the EOM are invariant, and can, therefore,
apply the argument from Sec.~\ref{sec:scale_inv} to deduce the scaling of the correlation functions (\ref{eq:corr_rho},\ref{eq:corr_g}). 

But why is it even possible to extract the scaling behavior from the coefficients at finite $k$, i.e. in a regulated theory with a finite IR cutoff? To see this, consider the
following argument (compare also to \cite{blaizot_pre06,chen_pre20}). Suppose, we have the inverse propagator in Fourier-space sufficiently close to the FP, i.e. $\Gamma_k^{(1,1,0,0)}(\omega,q)$ scales homogeneously under an RG transformation, $k \rightarrow s k$, and simultaneous rescaling of $q\rightarrow s q$ and $\omega\rightarrow s^z \omega$ (the scaling behavior of $\Gamma^{(1,1,0,0)}$ is inverse to that of $\bG$), i.e.
\beq
    \Gamma^{(1,1,0,0)}_k(\omega,q) = s^{\chi} \Gamma^{(1,1,0,0)}_{sk}(s^z \omega, s q),
\eeq
for some $\chi$. Now we can consider two equivalent cases. First we set $s= k^\prime/q$ with a constant scale $k^\prime$
\beq
    \Gamma^{(1,1,0,0)}_k(\omega,q) =
    \left(\frac{k^\prime}{q}\right)^{\chi} \Gamma^{(1,1,0,0)}_{k^\prime k/q}\left(\left(\frac{k^\prime}{q}\right)^z \omega, k^\prime\right) ,
\eeq
where we can now safely take the limit $k\rightarrow 0$, showing that the propagator of the effective action, which is the full solution to the many-body problem, all nonlinear fluctuations included, follows a powerlaw in $q$ with the exponent $\chi$.
Secondly we set $s= k^\prime/k$
\beqn
\nonumber
    \Gamma^{(1,1,0,0)}_k(\omega,q) &&= \left(\frac{k^\prime}{k}\right)^{\chi} \Gamma^{(1,1,0,0)}_{k^\prime}\left(\left(\frac{k^\prime}{k}\right)^z \omega, \frac{k^\prime  q}{k}\right) \\
    &&\approx \left(\frac{k^\prime}{k}\right)^{\chi} \alpha_0^\prime \ ,
\eeqn
where in the last line, we developed $\Gamma^{(1,1,0,0)}_k$ to zeroth order in $q$ and $\omega$, corresponding to our truncation which, by neglecting higher order derivative terms, also assumes that $q\ll k$. This last result shows, that the constant part of the inverse propagator scales in $k$ with the same exponent $\chi$. And this exponent has already  been obtained through our FRG analysis, i.e., $\chi=\eta_\mu-2$.

A similar argument can be made for the rest of the entries is $\Gamma_k^{(2)}$, showing the scaling behavior (\ref{eq:corr_rho},\ref{eq:corr_g}) for the realspace propagators, obtained from inverting and then Fourier-transforming $\Gamma_k^{(2)}$.

\begin{figure}
    \centering
    \includegraphics[width=\linewidth]{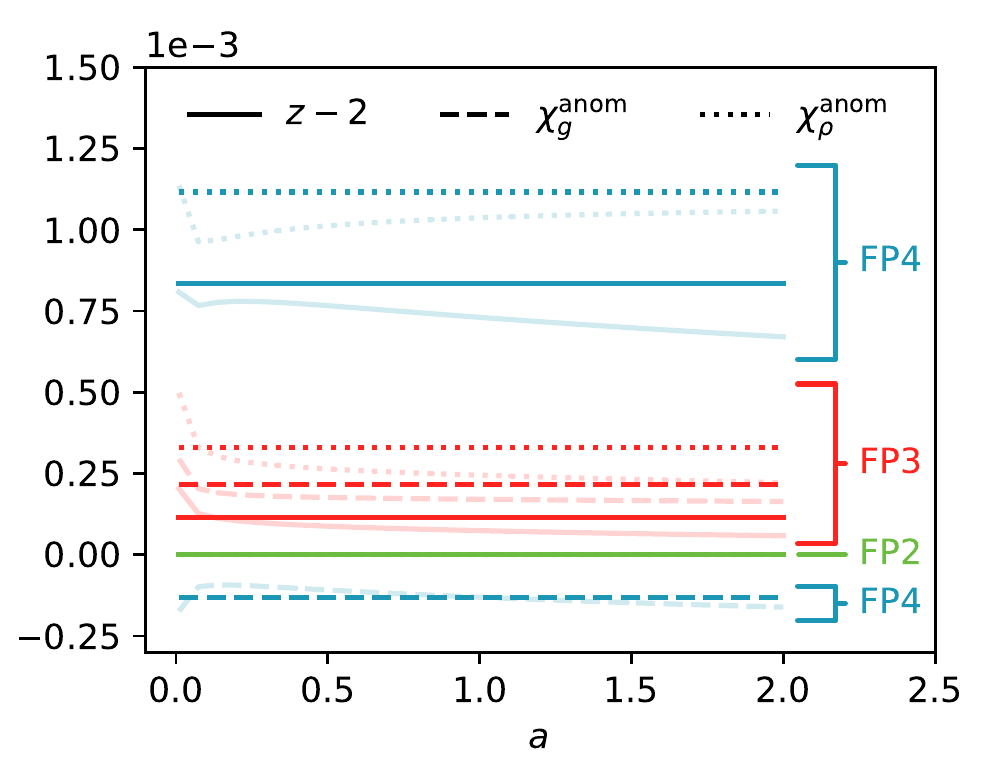}
    \caption{
    Dependence of the anomalous scaling exponents, $\chi_g^{\rm anom}=\chi_g+(d-2)/2$ and  $\chi_\rho^{\rm anom}=\chi_\rho+(d-4)/2$, on the regulator parameter $a$ obtained at $\epsilon=0.1$. The different colors label different fixed points, corresponding to Fig.~\ref{fig:RGflow}, and are annotated by brackets. The different line styles denote the different exponents. For the algebraic regulator (\ref{eq:alg_reg}), the exponents are completely independent of $a$ (saturated lines), which is why we accept them as our final results, see Tab.~\ref{tab}, according to the principle of minimal sensitivity. While the results obtained with the Litim regulator (\ref{eq:lit_reg}) do depend on $a$ (faint lines), they are compatible with those of the algebraic regulator and the deviations are always smaller than what is expected of next order corrections, i.e. smaller than $\epsilon^3=10^{-3}$.}
    \label{fig:pms}
\end{figure}

\subsection{Universal critical exponents and amplitude ratios}

The resulting values for the critical exponents at these FPs, depending on the regulator parameter $a$ are plotted in Fig.~\ref{fig:pms} for both regulators, Eq.~(\ref{eq:alg_reg}) and (\ref{eq:lit_reg}), exemplarily at $\epsilon=0.1$. This clearly shows that, for the algebraic regulator (\ref{eq:alg_reg}), the critical exponents are independent of the parameter $a$. While we have shown this numerically, an analytical argument for this is given in \cite{morris_plb94}. In contrast, the results using the Litim regulator (\ref{eq:lit_reg}) do depend on $a$, however, the estimated values for the critical exponents are compatible with those of the algebraic regulator, i.e. the deviations are all smaller than $\epsilon^3$. Due to the $\epsilon$ expansion accuracy can anyways only be expected to this degree. By virtue of the principle of minimal sensitivity \cite{balog_pre20}, we, therefore, chose the result of the algebraic regulator as our main results which, expressed in terms of $\epsilon = (d_c-d)$, are shown in Table \ref{tab}.

\setlength{\tabcolsep}{7pt}
	
	\begin{table*}
	 \caption{Universal critical exponents, expressed as an $\epsilon$-expansion from the upper critical dimension $d_c=6$, for the four distinct fixed points. 
	 }
		\label{tab}
\renewcommand{\arraystretch}{1.2}
        \begin{tabular}{cccccc}
        \hline
        \hline
        FP & $z-2$      & $\chi_g+(d-2)/2$ & $\chi_\rho+(d-4)/2$ & $y_1-2$ & $y_2-2$\\
        \hline
        1& $0$               &   $0$               & $0$                & $0$ & $0$ \\
        2& $0$               &   $0$               &   $0$ & $-0.33\epsilon$ & $0$\\
        3& $0.011\epsilon^2$ &   $0.022\epsilon^2$ & ${ 0.033\epsilon^2}$ & $-0.47\epsilon$ & $\mathcal O(\epsilon^2)$\\
        4& $0.084\epsilon^2$ &   ${ -0.013\epsilon^2}$ & ${ 0.112\epsilon^2}$ & $0.16\epsilon$ & $\mathcal O(\epsilon^2)$ \\
        \hline
        \hline
        \end{tabular}
    \end{table*}

In addition to these critical exponents, we can also provide quantitative predictions on the two universal amplitude ratios: $\bar \mu$ and $\bar \zeta$, shown in Tab.~\ref{tab:fps}. These could in principle be measured experimentally from diffusion constants and the wavelength of density waves.

Finally, we can also determine the exponents $y_1$ and $y_2$, describing the divergence of the correlation length according to Eq.~(\ref{eq:corr_length}), which are also shown in Tab.~\ref{tab}. 

The dependence of the amplitude ratios $\bar \mu$ and $\bar \zeta$, as well as the correlation length exponents $y_1$ and $y_2$ on the regulator parameter $a$ is similar to the other exponents in Fig.~\ref{fig:pms}, i.e. they are independent of $a$ for the algebraic regulator (\ref{eq:alg_reg}) and the results obtained with the Litim regulator (\ref{eq:lit_reg}), while varying in $a$, are compatible with those of the algebraic regulator up to a correction of next-to-leading order in $\epsilon$.

\subsection{FP2: a nonequilibrium version of the Ising universality class?}
\label{sec:fp2}
For the second fixed point, we observe that the inverse correlation length exponent $y_1$ agrees with the correlation length exponent of the Ising universality class in $4-\epsilon$ dimensions. Though the difference in upper critical dimension between the two universality classes implies their distinctness, it is nevertheless interesting to explore how this relationship arises, which we will do in this subsection. 

At FP2, the FP values of all $\alpha$-couplings vanish, which implies that the EOM reduce to
\begin{subequations}
\begin{align}
\partial_t\rho =&\ -\nabla \cdot \bg \\
\nonumber
\gamma \pp_t \bg=&\ \mu_1 \nabla^2 \bg +\mu_2\vnab (\vnab \cdot \bg)+\bff 
\\& 
 - \vnab \left( \kappa_0  \rho + \frac{\kappa_2}{3} \rho^3 -\zeta \nabla^2 \rho   \right) 
\ ,
\end{align}
\end{subequations}
which is linear in $\bg$. We can immediately see, that the transverse component decouples from both the density and longitudinal momentum field and that its dynamics are given by the mean field critical $O(d-1)$ EOM
\beq
\partial_t \bg^\bot = \mu_1 \nabla^2 \bg^\bot+\bff^\bot.
\eeq

In the parallel sector, we can eliminate $\bg^\parallel$ from the EOM to obtain
\beq
\label{eq:eomfp2}
(\gamma \partial_t - \mu_\parallel \nabla^2) \partial_t \rho = \nabla^2\left( \kappa_0 \rho + \frac{\kappa_2}{3} \rho^3 -\zeta \nabla^2 \rho \right) - \nabla \cdot \bff.
\eeq
This equation is again reminiscent of Model B dynamics, except that the time-derivative term is heavily modified. The linear mode of the momentum field, though eliminated from the equation, manifests now in the second order time-derivative. 
Here, all anomalous dimensions are zero. As a result, the fluctuation-dissipation relation is not explicitly broken and it remains to be seen whether the model equation (\ref{eq:eomfp2}) corresponds to an equilibrium system or not.

 This EOM can therefore be seen as an Ising model with exotic two-mode dynamics, that rises the scaling dimension of the field, and therefore also the upper critical dimension.

\subsection{FP4: Emergence of two time-scales}
\label{sec:two-timescales}

\setlength{\tabcolsep}{7pt}
	\begin{table*}
	 \caption{Fixed point values and universal critical exponents for the two additional fixed points in the $\bar\zeta=0$ plane, expressed as an $\epsilon$-expansion from the upper dimension $d_c=6$. The fixed point values of $\alpha_0$ and $\kappa_0$ are again normalized such that they are independent of the regulator parameter $a$.  When no value for $\bar\mu$ is given, it can take any arbitrary value and is not universal in this case.
	 }
		\label{tab3}

\begin{tabular}{ccccc || cc || cccc}
        \hline
        \hline
        FP & $\bar\alpha_0/a^{1/2}$ & $\bar\kappa_0/a^{1/2}$ & $\bar\alpha_2$ & $\bar\kappa_2$ &  $\bar\mu$ & $\bar \zeta$ & $\eta_\mu$  \rule{0pt}{3ex}     & $\eta_\gamma$ & $\eta_D$ & $\eta_\rho$   \\
        \hline
        2' &  $0$& $-0.399\epsilon$ & $0$ & $0.333\epsilon$ & & $0$& $0$               &   $0$               & $0$               & $0.044\epsilon^2$\\
        3' & ${ -0.261\epsilon}$& ${ -0.278\epsilon }$ &  ${ 0.217\epsilon}$ & ${ 0.231\epsilon}$ & $1.77$ & $0$ & $0.021\epsilon^2$ &   $0.052\epsilon^2$ & ${ 0.105\epsilon^2}$ & ${ 0.049\epsilon^2}$ \\
        \hline
        \hline
        \end{tabular}

\vspace{0.3cm}

        \begin{tabular}{cccccccc }
        \hline
        \hline
        FP \rule{0pt}{3ex} & $z-2$      & $\chi_g+(d-2)/2$ & $\chi_\rho+(d-4)/2$ & $y_1-2$ & $y_2-2$\\
        \hline
        2'& $0$               &   $0$               &   $-0.022\epsilon^2$ & $-0.33\epsilon$ & $0$\\
        3'& $0.031\epsilon^2$ &   $0.016\epsilon^2$ & ${ 0.022\epsilon^2}$ & $-0.45\epsilon$ & $\mathcal O(\epsilon^2)$ \\
        \hline
        \hline
        \end{tabular}

    \end{table*}

 So far, we have discussed the FPs, where $\eta_\rho = 0$. Now we turn to the case $\eta_\rho \neq 0$, which is the case for FP4. In this case, it is impossible to choose rescaling exponents $z$, $\chi_g$ and $\chi_\rho$ such that all the terms in the EOM (\ref{eq:nonlinscalingeom}) rescale homogeneously. For instance, if we were to choose the same exponents as for FP2 and 3 (\ref{eq:nlexponents}), in the large $\ell$, i.e. hydrodynamic limit, the continuity equation (\ref{eq:cont}) would reduce to a statement of staticality,
\begin{equation}
	\partial_t \rho = 0 \ ,
\end{equation}
and all the ``pressure-terms", proportional to $\nabla \rho$ in (\ref{eq:nonlinscalingeom}), would vanish, decoupling the momentum density field from the density field at the linear level. Since the continuity equation is modified, this is the only FP where the hyperscaling relation $\chi_\rho-\chi_g-1=z-2$, enforced by the extended symmetry of the continuity equation, is broken. 

If we interpret this as the momentum-density-couplings becoming irrelevant at this fixed point, we can simply omit these and obtain a scale invariance on the remaining terms. For the momentum density correlation function, we can therefore conclude that,
\begin{equation}
\bC_g(\br,t) =r^{2\chi_g} \bS_{gg}\left(\frac{t}{r^{z}}\right)\ .
\end{equation}
At the linear level, they would look just like in Eq.~(\ref{eq:corr_lin_g_para}) with $\zeta=0$. In this decoupled limit however, the density correlation function cannot be determined since the density field seemingly decouples from the noise term $\bff$.

This suggests, that this choice of rescaling is not the correct one when looking at density-density correlations. If we were to choose instead a different time and momentum field rescaling,
\begin{subequations}
\label{eq:altexponents}
\begin{align}
\label{eq:dynz}
&z^{\rm alt} = 2-\eta_\mu+\eta_\gamma -\eta_\rho \ , \\
\label{eq:chig}
&\chi_g^{\rm alt} = \frac{2-d-\eta_\mu-\eta_\gamma+\eta_D+\eta_\rho}{2}\ , \\
\label{eq:chirho}
&\chi_\rho^{\rm alt}= \chi_\rho  = \frac{4-d-3\eta_\mu+\eta_\gamma+\eta_D-{  \eta_\rho}}{2} \ .
\end{align}
\end{subequations}
the continuity equation remains scale invariant and the ``pressure-terms" stay relevant. However, now the time-derivative term on the left-hand-side of Eq.~(\ref{eq:nonlinscalingeom}) vanishes in the large $\ell$, i.e., hydrodynamic, limit.
Now, again regarding this term as irrelevant, the remaining terms support a scale invariance, from which we can conclude that the density correlation function scales as
\begin{equation}
C_\rho(\br,t) =r^{2\chi_\rho^{\rm alt}} S_{\rho\rho}\left(\frac{t}{r^{z^{\rm alt}}}\right) \ .
\end{equation}

At the linear level, with this rescaling, the correlation functions would be
\begin{subequations}
\begin{align}
    &
	\label{eq:corr_lin_rho}
    C_\rho(\br,t)=\frac{1}{\zeta}\int_{\tilde \bq} e^{\ii\tilde\bq\cdot\tilde \br} \frac{2Dq^2}{ q^8 +\omega^2 \mu_\parallel^2 q^4}
     \ , 
    \\
    &\bC_g^\parallel(\br,t)=\zeta \int_{\tilde \bq} e^{\ii\tilde\bq\cdot\tilde \br} \frac{2D \omega^2  \bP^\parallel(\bq)}{  q^8+\omega^2\mu_\parallel^2 q^4}  \ .
\end{align}
\end{subequations}
The transverse components of the momentum density remain unmodified.

This clearly shows that at FP4 two separate time-scales are emerging; one fast time-scale at which the momentum density can react quickly to perturbations in the presence of a density field frozen in time, and a slow time-scale over which the density relaxes once momentum fluctuations have already dissipated long ago. In other words, the momentum density ceases to be a hydrodynamic variable at this fixed point, since it becomes a ``fast" mode. 

Even though our analysis has shown that in the vicinity of FP4 the ``pressure-terms" proportional to $\zeta$ are irrelevant, they cannot be neglected in the RG analysis. First, as discussed in Sec.~\ref{sec:drg} if $\zeta$ is naively set to zero, the frequency integrals are clearly divergent. Secondly, the FP value of $\kappa_2$ plays an important role in determining the critical exponents of FP4. This coupling must therefore be included necessarily in the discussion.

The right coupling to neglect instead is therefore the time-derivative term of the momentum density. Then, its EOM becomes an exact, time-invariant identity enslaving the momentum field to the density field. It can be used to eliminate the momentum field from the continuity equation to obtain, for simplicity at the linear level, that
\begin{equation}
\left( \alpha_0-\mu_\parallel\nabla^2 \right) \partial_t \rho = \nabla^2 \left( \kappa_0 \rho -\zeta \nabla^2 \rho \right) -\nabla \cdot \bff \ ,
\end{equation}
which, except for the $\mu_\parallel$ term, is exactly the general equation of model B dynamics one would write down for a conserved, scalar quantity \cite{hohenberg_rmp77}. Fine-tuning $\kappa_0\rightarrow 0$ alone

yields the critical model B, Ising universality class. We can therefore interpret FP4 as a genuine nonequilibrium multicritical point of model B, where, in addition to $\kappa_0$, also the time-derivative term, characterized by $\alpha_0$, is fine-tuned to zero. For reasons of stability, the higher-order derivative term characterized by $\mu_\parallel$, is needed. 

The $\zeta=0$ manifold, therefore, describes the model B theory subspace. Within it, in addition to FP4, we find two additional FPs which are related to FP2 and FP3, so we name them FP2' and FP3' accordingly. Since they are unstable in the $\zeta$ direction and already described by the model B EOM, we will not  discuss them in detail. Instead, we will simply report here the FP values and critical exponents in Tab.~\ref{tab3}.

\section{Summary \& Outlook}
\label{sec:summary}

We have demonstrated in this work the whole process from first formulating the large scale hydrodynamic equations to discovering novel universality classes on the particular problem of the multicritical point in the phase diagram of compressible active matter, where the critical point of the flocking transition coincides with the critical point of disordered phase-separation.

We started with the Toner-Tu equations, derived purely from symmetry arguments and conservation laws, which therefore describe general compressible polar active fluids, i.e. systems of self-propelling and aligning particles at the hydrodynamic level. Using mean-field-theory and linear stability analysis one can show, first, the existence of two homogeneous phases and, secondly, the existence of regions where the homogeneous state is unstable indicating phase-separation, revealing the phase diagram of compressible polar active matter. Linear stability analysis further reveals the existence of two critical points, the critical order-disorder transition and the critical point of disordered phase-separation. Further, there is a multicritical point in the phase diagram where both these critical points overlap.

Analyzing the linearized equations of motion around this multicritical point, then revealed the scale invariance of the theory at the linear level, which results in powerlaw scaling of the correlation functions. We elucidated how these functions, and thereby the scaling exponents as well as the scaling behavior of the correlation length, could be measured in principle.

The scaling behavior of the linear theory further informed us about the relevance of the possible nonlinear terms and the critical dimension, above which the linear theory is expected to be exact. While in the nonlinear regime, the correlation functions could no longer be calculated explicitly, scale invariance will in general still lead to powerlaw correlation functions, though their scaling exponents are no longer trivially determined.

To determine these nontrivial exponents, we attempted a one-loop dynamic renormalization group (DRG) approach within the $\epsilon$-expansion, which ultimately failed, since two-loop effects are necessary to capture the universal physics of this problem.

To take these two-loop effects into account, we set up a functional renormalization group (FRG) ansatz which describes effective theories at intermediate scales and whose scale dependence is described by the Wetterich equation. A key ingredient to this approach was the physical insight that subsystems at an intermediate scale are locally in a homogeneous state even if the total system is phase separating. This enabled our FRG approach to go beyond the perturbative DRG approach.

Using various computer algebra and numerical methods developed for this work, the renormalization group flow equations could then be evaluated. Further, by inverting the sign of the flow equation for the relevant couplings, we found three renormalization group fixed points. Finding them proves the existence of {\it three} novel  universality classes, at least two of them being demonstrated to be genuinely out of equilibrium.
 
In summery, our achievements are three folds: (1) the discovery of three novel  universality classes, two of them being demonstrably out of  equilibrium \cite{ALP}, (2) the first analytical elucidation of critical behavior  for compressible active fluids, and (3) the first application of FRG on active matter systems beyond the equivalence of the perturbative one-loop level.

Interesting future directions include the applications of FRG to explore open questions in compressible active matter such as: what are the universality classes of the critical order-disorder transition \cite{nesbitt_njp21} and of the ordered phase \cite{toner_pre12,mahault_prl19}.

\newpage

\onecolumngrid

\begin{appendix}

\section{ FRG Flow equations}
\label{app:floweq}

In the following we give details on how the right-hand-side of Eqns.~(\ref{eq:fs}) can be obtained. As a first step, to reduce some of the complexity of the following calculation, we use a trick commonly used in FRG. We reformulate the Wetterich equation (\ref{eq:wetterich}) to the form
\begin{equation}
\label{eq:wetterich_tilde}
	\partial_\ell \Gamma_k = \frac{1}{2}  \left. \partial_{\ell^\prime} \tr\log(\Gamma_k^{(2)}+ R_{k^\prime})\right|_{k^\prime=k} \ ,
\end{equation}
where the derivative $\partial_{\ell^\prime}=k^{\prime}\partial_{k^\prime}$ on the right-hand-side, only acts on the $k$-dependence of the regulator. This reduces the number of propagators $(\Gamma_k^{(2)}+R_k)^{-1}$ in the following expressions by one.

As the projections in Eqns.~(\ref{eq:fs}) involve two functional derivatives of Eq.~(\ref{eq:wetterich_tilde}), the right-hand-side of Eqns.~(\ref{eq:fs}) after derivation will involve, in addition to $\Gamma_k^{(2)}$ also the third and fourth order functional derivatives of $\Gamma_k$:
\begin{equation}
\label{eq:wetterich2}
	\partial_\ell \Gamma_k^{(2)} = \partial_{\ell^\prime} \tr \left[ \frac{1}{2}(\Gamma_k^{(2)}+ R_{k^\prime})^{-1}\Gamma_k^{(4)}-(\Gamma_k^{(2)}+R_{k^\prime})^{-1}\Gamma_k^{(3)}(\Gamma_k^{(2)}+ R_{k^\prime})^{-1}\Gamma_k^{(3)} \right]_{k^\prime=k} \ ,
\end{equation}

where we didn't write explicitly the dependence of $\Gamma_k^{(3)}$ and $\Gamma_k^{(4)}$, on the external wavevectors  and frequencies that come from the functional derivatives. As written in Eqns.~(\ref{eq:fs}), the expressions are then to be evaluated at the constant field $\rho(\br,t) =\rho_{\rm unif}$. The expressions for $\Gamma_k^{(2)}$, $\Gamma_k^{(3)}$ and $\Gamma_k^{(4)}$ with this background field are straightforwardly determined from our ansatz, Eq.~(\ref{eq:G}):
\begin{subequations}
\label{eq:gammaderivs}
\begin{align}
    &\Gamma^{(0,0,1,1)}(\tilde \bq,\tilde \bp)=-\ii \omega_q\,\tilde \delta_{qp} \ , \\
    &\Gamma^{(0,1,1,0)}_i(\tilde \bq,\tilde \bp)=-\ii q_i\, \tilde\delta_{qp} \ , \\
    &\Gamma^{(2,0,0,0)}_{i,j}(\tilde \bq,\tilde \bp)=- 2 D \, \delta_{ij} \tilde\delta_{qp} \ , \\
    &\Gamma^{(1,1,0,0)}_{ij}(\tilde \bq,\tilde \bp)=(-\gamma\ii \omega_q \delta_{ij} +(\alpha_0+\alpha_2 \rho_{\rm unif}^2) \delta_{ij} +\mu_1 q^2 \delta_{ij} + \mu_2 q_i q_j )\tilde\delta_{qp}  \ , \\
    &\Gamma^{(1,0,0,1)}_i(\tilde \bq,\tilde \bp)= \ii (\kappa_0+\kappa_2 \rho_{\rm unif}^2+\zeta q^2) q_i \, \tilde\delta_{qp} \ , \\
    \label{eq:vert1}
    &\Gamma^{(1,1,0,1)}_{ij}(\tilde \bq,\tilde \bp,\tilde \bh)= 2\alpha_2 \rho_{\rm unif}\, \delta_{ij} \tilde\delta_{qph} \ ,\\
    &\Gamma^{(1,0,0,2)}_i(\tilde \bq,\tilde \bp,\tilde \bh)= 2 \ii \kappa_2 \rho_{\rm unif} q_i \,
    \tilde\delta_{qph} \ , \\
    &\Gamma^{(1,1,0,2)}_{ij}(\tilde \bq,\tilde \bp,\tilde \bh,\tilde \bu)= 2\alpha_2 \, \delta_{ij} \tilde\delta_{qphu} \ , \\
    \label{eq:vert4}
    &\Gamma^{(1,0,0,3)}_i(\tilde \bq,\tilde \bp,\tilde \bh,\tilde \bu)= 2 \ii \kappa_2 q_i \, \tilde\delta_{qphu},
\end{align}
\end{subequations}
where we introduced the short-hand notation $\tilde\delta_{qp\dots}=(2\pi)^{d+1}\delta^{d+1}(\tilde \bq+\tilde \bp+\dots)$. 
All other functional derivatives of $\Gamma_k$, up to fourth order, evaluated at these background fields and not mentioned here are vanishing.

From these equations (\ref{eq:gammaderivs}) we can determine the regulated propagator by inverting the $4\times 4$ matrix $\Gamma_k^{(2)}+R_k$,
%
\beq
\label{eq:propmatrix}
(\Gamma_k^{(2)}+R_k)^{-1}(\tilde\bq,\tilde \bp) = \begin{pmatrix}
        $0$ & \bG(-\tilde\bq) &$0$ & \bG(-\tilde\bq)\cdot \frac{\bq}{ \omega_q}  \\
        \bG(\tilde\bq) & \bG(\tilde\bq) 2D \bG(-\tilde\bq) & \ii\bG(\tilde\bq)\bm \cdot \frac{\bK(\tilde\bq)}{ \omega_q} & \bG(\tilde\bq) 2D \bG(-\tilde\bq) \cdot \frac{\bq}{ \omega_q} \\
        $0$ & -\ii \frac{\bK(-\tilde\bq)}{\omega_q} \cdot \bG(-\tilde\bq) &$0$ & H(-\tilde\bq) \\
        \frac{\bq}{ \omega_q} \cdot \bG(\tilde\bq) & \frac{\bq}{ \omega_q} \cdot  \bG(\tilde\bq) 2D \bG(-\tilde\bq) & H(\tilde\bq) & \frac{\bq}{ \omega_q} \cdot  \bG(\tilde\bq) 2D \bG(-\tilde\bq) \cdot \frac{\bq}{ \omega_q}
    \end{pmatrix}
    \tilde \delta_{qp} \,
\eeq
%
where we have defined,
%
\begin{subequations}
\label{eq:prop_frg}
\begin{align}
    &\bG(\tilde\bq)= \bP^\parallel(\bq) G^\parallel(\tilde\bq) + \bP^\bot(\bq) G^\bot(\tilde\bq) \ , \\
    & G^\parallel(\tilde\bq) = \frac{-\ii \omega_q }{-\ii \omega_q  \left(-\ii \gamma\omega_q+\alpha_0+\alpha_2\rho_{\rm unif}^2+A_{k^\prime}(q^2)+\mu_\parallel q^2 \right)+q^2 \left(\kappa_0+\kappa_2\rho_{\rm unif}^2+B_{k^\prime}(q^2)+\zeta  q^2\right)} \ , \\
    & G^\bot(\tilde\bq) = \frac{1}{-\ii \gamma\omega_q+\alpha_0+\alpha_2\rho_{\rm unif}^2+A_{k^\prime}(q^2)+\mu_1  q^2 } \ ,\\
    &H(\tilde q) = \frac{-\ii \gamma\omega_q+\alpha_0+\alpha_2\rho_{\rm unif}^2+A_{k^\prime}(q^2)+\mu_\parallel q^2 }{-\ii \omega_q  \left(-\ii \gamma\omega_q+\alpha_0+\alpha_2\rho_{\rm unif}^2+A_{k^\prime}(q^2)+\mu_\parallel q^2 \right)+q^2 \left(\kappa_0+\kappa_2\rho_{\rm unif}^2+B_{k^\prime}(q^2)+\zeta  q^2\right)} \ , \\
    & \bK(\tilde \bq) = \ii (\kappa_0+\kappa_2 \rho_{\rm unif}^2 +B_{k^\prime}(q^2)+\zeta q^2) \bq \ .
\end{align}
\end{subequations}
The propagator \bG \ is essentially the same as in the DRG calculation (\ref{eq:prop_drg}), except that it now contains the terms introduced by the regulator, which make the flow equations IR convergent, where in the DRG calculation this is taken care of by the integral boundaries.

Similarly to the DRG analysis (\ref{eq:feynman_rules}), we introduce a graphical notation
\begin{subequations}
\label{eq:feynman_rules_frg}
\begin{align}
&\diagram{28} = \bG(\tilde \bq) \ , \ \ \ \ 
\diagram{29} = \frac{1}{-\ii \omega_q} \ , \ \ \ \ 
\diagram{30} = -\ii \bq \ , \ \ \ \ 
\diagram{31} = 2D \ \id \ , \\
&\diagram{32} = \alpha_2\  \id \ , \ \ \ \ 
\diagram{33} = \ii \bq \kappa_2 \ ,\ \ \ \ 
\diagram{34} = \alpha_2 \rho_{\rm unif}\  \id \ , \ \ \ \ 
\diagram{35} = \ii \bq \kappa_2  \rho_{\rm unif}\ ,
\end{align}
\end{subequations}
which differs from the DRG notation only by the redefinition of the propagators (\ref{eq:prop_frg}) and the the three-point-vertices being nonzero.

If we regard the individual entries of Eq.~(\ref{eq:wetterich2}), especially those corresponding to the $\bF$'s in Eq.~(\ref{eq:fs}), we can arrange the vertices, Eqns.~(\ref{eq:vert1}-\ref{eq:vert4}), in matrices similar to that in Eq.~(\ref{eq:propmatrix}), such that the matrix products in Eq.~(\ref{eq:wetterich2}) can be carried out and traced over. Then
the $\bF$'s can be decomposed into individual terms, 
\begin{subequations}
\begin{align}
    \bF_g &= \id F_g + \bP^\parallel(\bar \bq) F_g^\parallel + \bP^\bot(\bar\bq) F_g^\bot \ , \\
    F_g^\parallel &= F_g^{\parallel, a} + F_g^{\parallel, b} +F_g^{\parallel, c} \ , \\
    F_g^\bot &= F_g^{\bot, a} + F_g^{\bot, b} \ , \\
    \bF_\rho &= \ii \bq \left[ F_\rho^{a} +  F_\rho^{b} + F_\rho^{c}+ F_\rho^{d} +F_\rho^{e} + F_\rho^{f}+ F_\rho^{g} \right] \ , \\
\bF_D &= \id \left[ F_D^a +F_D^b \right] \ ,
\end{align}
\end{subequations}
which can then be represented diagrammatically,
\begin{subequations}
\begin{align}
&\id F_g = \diagram{2} , &&  F_g^{\parallel, a}+ F_g^{\bot, a}= -4 \diagram{3}, && F_g^{\parallel, b}+ F_g^{\bot, b}= -4\diagram{4},\\
&F_g^{\parallel, c}= -4\diagram{5},&&\ii \bq  F_\rho^{a}= \diagram{6}, &&\ii \bq  F_\rho^{b}=-4\diagram{7},\\
&\ii \bq  F_\rho^{c}=-4\diagram{8},&&\ii \bq  F_\rho^{d}=-4\diagram{9},&&\ii \bq  F_\rho^{e}=-4\diagram{10},\\ &\ii \bq  F_\rho^{f}=-4\diagram{11},&&\ii \bq  F_\rho^{g}=-4\diagram{12}, && \id  F_D^a = -4 \diagram{13}, \\ 
& \id  F_D^b = -4 \diagram{14}.
\end{align}
\end{subequations}
In dimensionless units, 
\begin{subequations}
\begin{align}
    &\bar G^\parallel = \mu_\parallel k^2 G^\parallel \ , \\
    &\bar G^\bot = \mu_\parallel k^2 G^\bot \ ,
\end{align}
\end{subequations}
their analytical expressions are given by
%
\begin{subequations}
\label{eq:fsdetailed}
\begin{align}
    &\bar F_g = \int^\prime_{\tilde \bp} 2 \bar \alpha_2 {\bar\zeta} \bar p^2 \frac{\bar G^\parallel(\tilde\bp)\bar G^\parallel(-\tilde\bp)}{\bar \omega_p^2} \ , \\
    &\bar F_g^{\parallel,a} = - \int^\prime_{\tilde \bp}  8 \bar \alpha_2^2\bar \rho_{\rm unif}^2 {\bar\zeta} (\bar q-\bar pz)\bar pz \frac{\bar G^\parallel(\tilde\bq-\tilde\bp)}{-\ii(\bar \omega_q-\bar \omega_p)}\frac{\bar G^\parallel(\tilde\bp)\bar G^\parallel(-\tilde\bp) }{\ii \bar\omega_p} \ , \\
    &\bar F_g^{\bot,a} = - \int^\prime_{\tilde \bp}  8 \bar \alpha_2^2\bar \rho_{\rm unif}^2{\bar\zeta} \bar p^2(z^2-1) \frac{\bar G^\parallel(\tilde\bq-\tilde\bp)}{-\ii(\bar \omega_q-\bar \omega_p)}\frac{\bar G^\parallel(\tilde\bp)\bar G^\parallel(-\tilde\bp) }{\ii \bar\omega_p}  \ , \\
    &\bar F_g^{\parallel,b} = - \int^\prime_{\tilde \bp}  8 \bar \alpha_2^2\bar \rho_{\rm unif}^2{\bar\zeta} \bar p^2\frac{\bar p^2(1-z^2)\bar G^\bot(\tilde\bq-\tilde\bp) + (\bar q-\bar p z)^2\bar G^\parallel(\tilde\bq-\tilde\bp)}{\bar p^2+\bar q^2-2\bar p\bar q z}  \frac{\bar G^\parallel(\tilde\bp)\bar G^\parallel(-\tilde\bp)}{\bar \omega_p^2} \ , \\
    \nonumber
    &\bar F_g^{\bot,b} = - \int^\prime_{\tilde \bp}  8 \bar \alpha_2^2\bar \rho_{\rm unif}^2{\bar\zeta} \bar p^2\left[\left(d-1-\frac{\bar p^2(1-z^2)}{\bar p^2+\bar q^2-2\bar p\bar q z}\right) \bar G^\bot(\tilde\bq-\tilde\bp) + \left(1- \frac{(\bar q-\bar p z)^2}{\bar p^2+\bar q^2-2\bar p\bar q z} \right)\bar G^\parallel(\tilde\bq-\tilde\bp) \right] \\
    &\hspace{4cm}\times\frac{\bar G^\parallel(\tilde\bp)\bar G^\parallel(-\tilde\bp)}{\bar \omega_p^2} \ , \\
    &\bar F_g^{\parallel,c} = - \int^\prime_{\tilde \bp}  8 \bar \alpha_2 \bar \kappa_2 \bar \rho_{\rm unif}^2{\bar\zeta^2} (\bar q^2-\bar q\bar p z) \frac{\bar G^\parallel(\tilde\bq-\tilde\bp)}{-\ii(\bar \omega_q-\bar \omega_p)} \frac{\bar G^\parallel(\tilde\bp)\bar G^\parallel(-\tilde\bp)}{\bar \omega_p^2} \ , \\
    &\bar F_\rho^a = \int^\prime_{\tilde \bp} 2 \bar \kappa_2{\bar\zeta} \bar p^2 \frac{\bar G^\parallel(\tilde\bp)\bar G^\parallel(-\tilde\bp)}{\bar \omega_p^2} \ , \\
    &\bar F_\rho^{b} = \int^\prime_{\tilde \bp}  8 \bar \alpha_2^2 \bar \rho_{\rm unif}^2 \frac{\bar G^\parallel(\tilde\bq-\tilde\bp)}{-\ii(\bar \omega_q-\bar \omega_p)} \left[ \frac{(\bar qz-\bar p)z}{\bar q} \bar G^\parallel(\tilde\bp)\bar G^\parallel(-\tilde\bp) +(1-z^2) \bar G^\bot(\tilde\bp)\bar G^\bot(-\tilde\bp)\right] \ , \\
    &\bar F_\rho^{c} = \int^\prime_{\tilde \bp}  8 \bar \alpha_2^2 \bar \rho_{\rm unif}^2 \left[ \frac{(\bar q\bar pz-\bar p^2)(\bar q-\bar p z)}{(\bar p^2+\bar q^2-2\bar p\bar q z)\bar q} \bar G^\parallel(\tilde\bq-\tilde\bp) + \left(\frac{\bar p z}{\bar q}-\frac{(\bar q\bar pz-\bar p^2)(\bar q-\bar p z)}{(\bar p^2+\bar q^2-2\bar p\bar q z)\bar q} \right) \bar G^\bot(\tilde\bq-\tilde\bp) \right] \frac{\bar G^\parallel(\tilde\bp) \bar G^\parallel(-\tilde\bp)}{-\ii\bar \omega_p}\ , \\
    &\bar F_\rho^{d} = -\int^\prime_{\tilde \bp}  8 \bar \alpha_2 \bar \kappa_2 \bar \rho_{\rm unif}^2{\bar\zeta}  (\bar p^2-\bar p\bar q z) \frac{\bar G^\parallel(\tilde\bq-\tilde\bp)}{-\ii(\bar \omega_q-\bar \omega_p)} \frac{\bar G^\parallel(\tilde\bp) \bar G^\parallel(-\tilde\bp)}{-\ii\bar \omega_p}\ , \\
    &\bar F_\rho^{e} = -\int^\prime_{\tilde \bp}  8 \bar \alpha_2 \bar \kappa_2 \bar \rho_{\rm unif}^2{\bar\zeta}  \frac{\bar p z(\bar p^2+\bar q^2-2\bar p\bar q z)}{\bar q} \frac{\bar G^\parallel(\tilde\bq-\tilde\bp)}{-\ii(\bar \omega_q-\bar \omega_p)} \frac{\bar G^\parallel(\tilde\bp) \bar G^\parallel(-\tilde\bp)}{\ii\bar \omega_p}\ , \\
    &\bar F_\rho^{f} = -\int^\prime_{\tilde \bp}  8 \bar \alpha_2 \bar \kappa_2 \bar \rho_{\rm unif}^2{\bar\zeta}  \frac{\bar p^2(\bar q-\bar p z)}{\bar q} \bar G^\parallel(\tilde\bq-\tilde\bp) \frac{\bar G^\parallel(\tilde\bp) \bar G^\parallel(-\tilde\bp)}{\bar \omega_p^2}\ , \\
    &\bar F_\rho^{g} = -\int^\prime_{\tilde \bp}  8 \bar \kappa_2^2 \bar \rho_{\rm unif}^2{\bar\zeta^2} (\bar p^2+\bar q^2-2\bar p\bar q z) \frac{\bar G^\parallel(\tilde\bq-\tilde\bp)}{-\ii(\bar\omega_q-\bar\omega_p)} \frac{\bar G^\parallel(\tilde\bp) \bar G^\parallel(-\tilde\bp)}{\bar \omega_p^2}\ , \\
&\bar F_D^a = \int^\prime_{\tilde \bp}16 \bar\alpha_2^2 \bar\rho_{\rm unif}^2 \bar\zeta \bar p^2 \left( \frac{\bar G^\parallel(\tilde\bp)\bar G^\parallel(-\tilde\bp)}{-\ii\bar\omega_p} \right)^2\ , \\
&\bar F_D^b = -\int^\prime_{\tilde \bp}16 \bar\alpha_2^2 \bar\rho_{\rm unif}^2 \bar\zeta \bar p^2 \left( \bar G^\parallel(\tilde\bp)\bar G^\parallel(-\tilde\bp)+\bar G^\bot(\tilde\bp)\bar G^\bot(-\tilde\bp)  \right) \frac{\bar G^\parallel(\tilde\bp)\bar G^\parallel(-\tilde\bp)}{\bar\omega_p^2}   \ ,
\end{align}
\end{subequations}
%
where we introduced the cosine between the internal and external wave vector $z = \bar\bq\cdot\bar\bp/\bar q\bar p$ and the ``primed integral",
\beq
\label{eq:intprime}
    \int^\prime_{\tilde \bp} \dots = \left. \int \frac{\dd{^d\bar p}}{(2\pi)^d} \int \frac{\dd{\bar \omega_p}}{2\pi}\partial_{\ell^\prime} \dots \right|_{k^\prime = k}.
\eeq
Other terms, not listed in Eq.~(\ref{eq:fsdetailed}), one would naively expect to appear, but vanish for reasons of causility (see e.g. \cite{canet_jopa11}), i.e., terms that have poles only in the lower/upper half of the complex plane in the variable $\bar \omega_p$.
Notice that the limit $\epsilon\ll 1$ has not yet been applied yet.

From this point on, further calculations by hand quickly become unmanagable, even though analytical treatment is still possible for a few more steps. Going from here to integrating the flow equations (\ref{eq:frg_flow}) is described in App.~\ref{app:numerics}.

\section{Computer algebra and numerical methods}
\label{app:numerics}

In the following, we describe how, from the equations above (\ref{eq:frg_flow}, \ref{eq:fsdetailed}), one can use computer algebra and numerical methods to solve the flow equations to find FP solutions and obtain the corresponding critical exponents characterizing their universality class.

For each $F$ in Eq.~(\ref{eq:fsdetailed}), one has to solve the three different integrals as defined in Eq.~(\ref{eq:intprime}): the frequency integral over $\bar \omega_p$, the angular integral over the directional degrees of freedom of $\bar \bp$ and the integral over the wavenumber $\bar p$. The frequency integrals can be solved exactly using Cauchy's integral formula and the angular integral can be solved analytically in the limit of $\bar q \ll 1$, which fortunately is the limit we are interested in. However, the remaining integral over $\bar p$ must be solved numerically since $\bar \mu$ and $\bar \zeta$ are nonnegligible, scale-dependent numbers.

Additionally, one has to perform the $\partial_{\ell^\prime}$ derivative in (\ref{eq:intprime}), the expansion into small $\epsilon$ and the expansion into small external wave numbers and frequencies, $\bar q$ and $\bar \omega_q$, (depending on which couplings, one either needs the zeroth or second order Taylor coefficient in $\bar q$ and zeroth or first order Taylor coefficient in $\bar \omega_q$ according to Eq.~(\ref{eq:frg_flow})). All three of these operations are applied after the frequency integral, where it is most convenient.

As the pole-structure in any of the integrands of Eqns.~(\ref{eq:fsdetailed}) is simple enough - one can chose to close the contour such that the only possible poles appearing are 
%
\beq
\label{eq:long_pole}
    \bar \omega_p = \pm \frac{1}{2} \ii \left(\bar p^2+\bar\alpha_0+\bar \alpha_2\bar \rho_{\rm unif}^2+\bar A_{k^\prime}(\bar p^2) \pm \sqrt{\left(\bar p^2+\bar\alpha_0+\bar \alpha_2\bar \rho_{\rm unif}^2+\bar A_{k^\prime}(\bar p^2)\right)^2-4 \bar \zeta \bar p^2 \left(  \bar p^2 + \bar\kappa_0 + \bar\kappa_0 \bar \rho_{\rm unif}^2+ \bar B_{k^\prime}(\bar p^2)\right)}\right) , 
\eeq
%
and
\begin{equation}
    \bar \omega_p = \pm \ii \left(\bar \mu \bar p^2+\bar A_{k^\prime}(\bar p^2)\right),
\end{equation}
i.e., independent of the external wave vector $\bar q$, and where we have defined $\bar A_{k^\prime}(\bar p^2) = A_{k^\prime}(\bar p^2)/[\mu_\parallel k^2]$ and $\bar B_{k^\prime}(\bar p^2) = B_{k^\prime}(\bar p^2)/[\zeta k^2]$ - one can apply Cauchy's integral formula to solve the frequency integrals analytically.

Then the $\partial_{\ell^\prime}$ derivative can be carried out, which is straightforward since only $\bar A_{k^\prime}$ and $\bar B_{k^\prime}$ depend on $k^\prime$ and their derivatives evaluate to
\begin{subequations}
\label{eq:regderivs}
\begin{align}
    \left. \bar A_{k^\prime}(\bar q^2) \right|_{k} &= m(\bar q^2) \ , \\
     \left. \partial_{\ell^\prime} \bar A_{k^\prime}(\bar q^2) \right|_{k} &= (\eta_\mu-2) m(\bar q^2) + 2 \bar q^2 m^\prime(\bar q^2) \ , \\
     \left. \bar B_{k^\prime}(\bar q^2) \right|_{k} &= m(\bar q^2) \ , \\
     \left. \partial_{\ell^\prime} \bar B_{k^\prime}(\bar q^2) \right|_{k} 
     &= (2\eta_\mu+\eta_\rho-\eta_\gamma  -2) m(\bar q^2) + 2 \bar q^2 m^\prime(\bar q^2) \ . 
\end{align}
\end{subequations}
Keep in mind that $\bar A_{k^\prime}$ and $\bar B_{k^\prime}$ can appear with $\bar q^2$ or $|\bar \bp-\bar \bq|^2$ as arguments. Since the anomalous dimensions $\eta_\mu$, $\eta_\rho$ and $\eta_\gamma$ are of $\mathcal O(\epsilon^2)$, they can be neglected in Eqns.~(\ref{eq:regderivs}). This is for example a convenient point to perform the expansion in small $\epsilon$, noting that the couplings $\bar \alpha_0$, $\bar \kappa_0$, $\bar \alpha_2$ and $\bar \kappa_2$ are of order $\mathcal O(\epsilon)$.

Finally, as a preparatory step for the angular integrals, the resulting integrands are developed in small $\bar q$ (to second order) and $\bar \omega_q$ (to first order). This produces derivatives of $m$ up to fourth order.

The only angular dependence of the integrals comes from the dependence of $z$ which, due to Taylor expansion into small $\bar q$, is polynomial up to order $z^4$. Using the Jacobian for $d$-dimensional spherical coordinates, one can show that
\begin{equation}
\label{eq:angint}
    \int \frac{\dd{^d \bar p}}{(2\pi)^d} =
    \frac{\Gamma(\frac{d-1}{2})}{\sqrt \pi \Gamma(\frac{d}{2})}\frac{S_d}{(2\pi)^d} \int_0^\infty \dd{ \bar p}\, \bar p^{d-1} \int_{-1}^1 \dd z (1-z^2)^\frac{d-3}{2} =\int_0^\infty \dd{\bar p}\,  \bar p^{d-1} \int_z \ .
\end{equation}
 An integral over any power of $z$ and, therefore, also any polynomial of $z$, can be performed using the formula
\begin{equation}
    \int_z z^n=\frac{\Gamma(\frac{d-1}{2})}{\sqrt \pi \Gamma(\frac{d}{2})} \int_{-1}^1 \dd z (1-z^2)^\frac{d-3}{2} z^{n} = \frac{\Gamma(\frac{d}{2})\Gamma(\frac{1+n}{2})}{2\sqrt\pi \Gamma(\frac{d+n}{2})} \ ,
\end{equation}
for even $n$. For odd $n$ the angular integral vanishes.

Finally, the wave number integral can be reformulated in terms of $y=\bar p^2$,
\beq
\label{eq:wnint}
    \int_0^\infty \dd{\bar p}\, \bar p^{d-1} = \frac{1}{2}\int_0^\infty \dd{y}\, y^{(d-1)/2}.
\eeq

\begin{algorithm}
\DontPrintSemicolon
\SetKwProg{feq}{Calculate flow equation}{:}{\KwRet}
\caption{The numerical procedure to obtain the FP solutions and critical exponents}\label{alg}
Select FP\;
Select Regulator type and parameter\;
$\bar \mu \gets \bar \zeta \gets 1$\;
$\bar \alpha_0 \gets \bar \kappa_0 \gets 0$\;
$\bar \alpha_2 \gets \bar \kappa_2 \gets 10^{-2}\epsilon$\;
Invert sign of $\alpha_0$ and $\kappa_0$ flow equation\;
\uIf{FP2}{
    $\alpha_2\gets 0$\;
}
\ElseIf{FP3}{
    $\kappa_2\gets-\kappa_2$\;
    Invert sign of $\kappa_2$ flow equation\;
}
$\ell \gets 0$\;
$\dd \ell\gets 5\times 10^{-4}$\;
Initialize GSL ODE solver environment with standard control, rfk45 step type and errors of ${\rm epsabs} \gets 10^{-11}$, ${\rm epsrel} \gets 10^{-13}$\;
\While{$\ell < 500/\epsilon^2$}{ 
    \feq{$(\bar\mu,\bar\zeta,\bar\alpha_2,\bar\kappa_2,\bar\alpha_0,\bar\kappa_0)$}{
        \uIf{algebraic regulator}{
            Initialize GSL QAG integral environment with {\it infinite} boundaries and errors of ${\rm epsabs} \gets 10^{-12}$, ${\rm epsrel} \gets 10^{-10}$\;
        }
        \ElseIf{Litim regulator}{
            Initialize GSL QAG integral environment with {\it finite} boundaries, 61 point Gauss-Kronrod rule and errors of ${\rm epsabs} \gets 10^{-12}$, ${\rm epsrel} \gets 10^{-10}$\;
        }
        Calculate integrals in flow equation\;
        Calculate flow equation\;
        \KwRet{value of flow equation} 
    }
    Perform ODE step\;
}
Restore original signs of flow equations\;
Take a discrete  derivative with finite difference $\dd c=10^{-7}$ of {\bf flow equation} with respect to all couplings\;
Calculate 2 largest eigenvalues of the derivative matrix\;
Save FP values and 2 largest eigenvalues\;
\end{algorithm}

All steps in App.~\ref{app:numerics} so far have been performed using computer algebra. The resulting expressions have then been converted into C++ code such that the final wave number integral can be solved numerically. The complete set of RG flow equations (\ref{eq:frg_flow}) is then solved using a fourth order  adaptative Runge-Kutta-Fehlberg (4,5) algorithm  provided by the GNU Scientific Library \cite{galassi_b09}, where at each RG ``time-step", the wave number integral (\ref{eq:wnint}) in (\ref{eq:fsdetailed}) is solved numerically using an adaptative quadrature routine, with the 15 point Gauss-Kronrod rule for infinite boundary integrals in case of the algebraic regulator and the 61 point Gauss-Kronrod rule for finite boundary integrals in case of the Litim regulator, again provided by the GNU Scientific Library \cite{galassi_b09}. 
For the adaptative ODE-solver we set a maximum relative and absolute error of $e_r = 10^{-11}$ and $e_a = 10^{-13}$ respectively. For the integration they are set to $e_r = 10^{-12}$ and $e_a = 10^{-10}$. 

Since, we are modifying the flow equations such that each fixed point we would like to investigate is attractive, the initial conditions do not matter too much (as long as they are within the attractive basin). In our analysis we set them close to the Gaussian FP. Since we obtained all our results in Tab.~\ref{tab} at $\epsilon=0.1$, the following initial conditions are sufficiently close to the Gaussian FP: $\bar\mu=\bar \zeta=1$, $\bar\alpha_2=\bar\kappa_2=10^{-3}$ and $\bar\alpha_0=\bar\kappa_0 = 0$. We use these initial conditions to analyze all FPs, except for the following modifications: if we are interested in FP2, we set $\bar\alpha_2=0$ (with this condition it is not necessary to invert the flow equation for $\bar\alpha_2$, as it is not being generated if it is vanishing initially), and when we investigate FP4, we set $\bar\kappa_2=-10^{-3}$.

When solving the ODEs we set an initial ``time-step" of $\dd \ell = 5 \times 10^{-4}$ initially, but it is being quickly changed by the adaptative algorithm. The adaptative nature of the algorithm is key to solve this problem, since there are three vastly different RG ``time-scales", given by the critical exponents, to resolve. First, the relevant couplings $\alpha_0$ and $\kappa_0$ converge very quickly with a time-scale of $\ell\sim 0.5$ to the critical surface. The next time-scale is that of nonlinear couplings $\alpha_2$ and $\kappa_2$ and of order $\ell \sim 1/\epsilon$. Then the final and longest time-scale sets in, wherein the amplitude ratios $\bar\mu$ and $\bar\zeta$ converge. This time-scale is approximately of the size of the anomalous dimensions, i.e., of order $\ell\sim 1/\epsilon^2$. However, since the prefactor of the $\mathcal O(\epsilon^2)$ critical exponents is typically a lot smaller than unity, we find total convergence of the fixed points only at about $\ell\sim 5\times 10^{4}$, though FP4 converges somewhat quicker, since the scaling exponents are generally larger, compare Fig.~\ref{fig:pms} and Tab.~\ref{tab}.

Finally, once the ODEs have converged to the desired FP, the correlation length exponents can be obtained by first restoring the original signs of the flow equations, and then taking discrete derivatives with a finite difference of $\dd c=10^{-7}$ with respect to the couplings. The largest two eigenvalues of the so obtained matrix characterizing the linearized flow equations around the FP give the correlation length exponents $y_1$ and $y_2$. 

The complete algorithm is summarized again in Alg.~\ref{alg}.

\section{DRG Flow equations}
\label{app:drg}

Instead of going through the usual procedure of obtaining the graphical corrections to the DRG flow equations, we instead obtain them from the FRG calculation. The flow equations are still described by Eqns.~(\ref{eq:frg_flow},\ref{eq:anom_dim_flow}), but the $\bF$'s in Eq.~(\ref{eq:fsdetailed}) need to be adapted. While there is no finite background field in the DRG formalism, it is nevertheless important to realize, that the derivatives with respect to $\rho_{\rm unif}$ in Eq.~(\ref{eq:frg_graphical}) essentially represent attachments of external legs with vanishing wavevector to the diagrams, e.g.,
\begin{equation}
\frac{\partial^2}{\partial \bar\rho_{\rm unif}^2} \diagram{7} = 2 \diagram{22} \ .
\end{equation}
This important fact creates the relationship between FRG diagrams and the corresponding DRG diagrams. Only after these derivatives have been performed, the background field is set to $\rho_{\rm unif}=0$ in all expressions, since in the perturbative DRG the background field is vanishing. Since no derivatives with respect to $\rho_{\rm unif}$ appear in the projections for the higher order derivative term couplings, i.e. $\mu_1$, $\mu_\parallel$ and $\zeta$, their graphical corrections are therefore directly set to zero since they are proportional to $\rho_{\rm unif}$.

Secondly, the regulator needs to be redefined to a sharp cutoff. The technical aspects of this cutoff are well described in \cite{morris_NPhB96}, but here we will describe its practical implications: 1) Any regulator functions in the integrands can directly be set to zero, $\bar A_{k^\prime}=\bar B_{k^\prime}=0$. 2) The frequency integral remains unmodified. 3) In diagrams where the external wavevector is nonzero, the integrand needs to be averaged over the two possible paths the external wavevector can take through the diagram, i.e. $1/2$ the integrands from Eqn. (\ref{eq:fsdetailed}) plus $1/2$ the same integrand with the replacement $\bar \bp \rightarrow \bar \bp+\bar\bq$. This step can be skipped in the small $\epsilon$ expansion since the graphical corrections are independent of the wavevector routing to linear order in $\epsilon$. 4) The integral over the magnitude of the loop wavevector and the $\partial_{\ell^\prime}$ derivative are removed, i.e. 
\begin{equation}
\int_0^\infty\dd\bar p \partial_{\ell^\prime}\rightarrow 1.
\end{equation}
5) The magnitude of the dimensionless wavenumber is set to unity $\bar p\rightarrow 1$.

These five rules can formally be derived by realizing that the sharp regulator amounts to a change in regulator:
\begin{equation}
\bG(\tilde \bq) \rightarrow \left.\bG(\tilde \bq) \Theta(k^\prime-\bq) \right|_{A_{k^\prime}=B_{k^\prime}=0}.
\end{equation}
The theta and delta functions appearing after taking the $\partial_{\ell^\prime}$ derivative realize the rules described above. The DRG flow equations obtained this way have been displayed in Eq.~(\ref{eq:drg_flow}).

As an example consider the term $F_\rho^d$. In the full FRG formalism, it would contribute to the graphical corrections of $\kappa_2$ and $\zeta$. But because it is proportional to $\rho_{\rm unif}^2$, it can only contribute to $\kappa_2$ in the adapted DRG equations, since it is the only coupling requiring two derivatives with respect to $\rho_{\rm unif}$ in its projection according to Eq.~(\ref{eq:frg_graphical}). According to the rules outlined above we then have to make the following replacements to obtain the contribution to $\kappa_2$ ($\alpha_0=\kappa_0=0$ is assumed in the following expression in accordance with the small $\epsilon$ expansion to simplify the expressions):
\begin{subequations}
\begin{align}
&\left.\frac{\partial^2}{\partial \bar\rho_{\rm unif}^2} F_\rho^{d}\right|_{\rho_{\rm unif}=0}\\
 &\hspace{0.5cm}= -\int^\prime_{\tilde \bp}  16 \bar \alpha_2 \bar \kappa_2 {\bar\zeta}  (\bar p^2-\bar p\bar q z) \frac{\bar G^\parallel(\tilde\bq-\tilde\bp)}{-\ii(\bar \omega_q-\bar \omega_p)} \frac{\bar G^\parallel(\tilde\bp) \bar G^\parallel(-\tilde\bp)}{-\ii\bar \omega_p}\ , \\
&\hspace{0.5cm}\xrightarrow{1),2)} -\int^\prime_{\tilde \bp}  \frac{\ii \bar\omega_p 16 \bar \alpha_2 \bar \kappa_2 {\bar\zeta}  (\bar p^2-\bar p\bar q z)}{-\ii  (\bar\omega_q-\bar\omega_p) \left(-\ii (\bar\omega_q-\bar\omega_p)+|\bar \bq-\bar \bp|^2 \right)+ \bar\zeta |\bar \bq-\bar \bp|^4}  \left|\frac{1 }{-\ii \bar \omega_p \left(-\ii \bar\omega_p+\bar p^2 \right)+ \bar\zeta  \bar p^4}\right|^2 \ , \\
\label{eq:twopaths}
\nonumber
&\hspace{0.5cm}\xrightarrow{3),4)} -\int_z \int \frac{\dd \bar \omega_p}{2\pi}\frac{\bar p^{d-1}}{2} \left\{ \frac{ \ii \bar\omega_p 16 \bar \alpha_2 \bar \kappa_2 {\bar\zeta}  (\bar p^2-\bar p\bar q z)}{-\ii  (\bar\omega_q-\bar\omega_p) \left(-\ii (\bar\omega_q-\bar\omega_p)+|\bar \bq-\bar \bp|^2 \right)+ \bar\zeta |\bar \bq-\bar \bp|^4}  \left|\frac{1 }{-\ii \bar \omega_p \left(-\ii \bar\omega_p+\bar p^2 \right)+ \bar\zeta  \bar p^4}\right|^2 \right.  \\
&\hspace{1.5cm}\left. +\frac{  \ii \bar\omega_p 16 \bar \alpha_2 \bar \kappa_2 {\bar\zeta}  (\bar p^2+\bar p\bar q z)}{-\ii  (\bar\omega_q-\bar\omega_p) \left(-\ii (\bar\omega_q-\bar\omega_p)+p^2 \right)+ \bar\zeta p^4}  \left|\frac{1 }{-\ii \bar \omega_p \left(-\ii \bar\omega_p+ |\bar \bq+\bar \bp|^2 \right)+ \bar\zeta   |\bar \bq+\bar \bp|^4}\right|^2 \right\} \ , \\
\nonumber
&\hspace{0.5cm}\xrightarrow{5)} -\int_z \int \frac{\dd \bar \omega_p}{2\pi}\frac{1}{2} \left\{ \frac{ \ii \bar\omega_p 16 \bar \alpha_2 \bar \kappa_2 {\bar\zeta}  (1-\bar q z)}{-\ii  (\bar\omega_q-\bar\omega_p) \left(-\ii (\bar\omega_q-\bar\omega_p)+(1+\bar q^2-2\bar q z) \right)+ \bar\zeta (1+\bar q^2-2\bar q z)^2}  \left|\frac{1 }{-\ii \bar \omega_p \left(-\ii \bar\omega_p+1 \right)+ \bar\zeta  }\right|^2 \right.  \\
&\hspace{1.5cm}\left. +\frac{  \ii \bar\omega_p 16 \bar \alpha_2 \bar \kappa_2 {\bar\zeta}  (1+\bar q z)}{-\ii  (\bar\omega_q-\bar\omega_p) \left(-\ii (\bar\omega_q-\bar\omega_p)+1 \right)+ \bar\zeta}  \left|\frac{1 }{-\ii \bar \omega_p \left(-\ii \bar\omega_p+ (1+\bar q^2+2\bar q z) \right)+ \bar\zeta  (1+\bar q^2+2\bar q z)^2}\right|^2 \right\} \ , 
\end{align}
\end{subequations}
where the angular integral $\int_z$ is defined according to Eq.~(\ref{eq:angint}). Note that the wavevector $\bar\bq$ and the cosine $z$ are technically different physical objects in the first and second term of Eq.~(\ref{eq:twopaths}), but have been relabeled after the variable transformation.

\end{appendix}

\twocolumngrid

\bibliography{references}

\end{document}